\documentclass[sigplan,10pt,screen,nonacm]{acmart}
\usepackage{paralist}
\usepackage{multirow}
\usepackage{xspace,wrapfig}
\usepackage[subtle,leadingfraction=1,charwidthfraction=1]{savetrees}
\usepackage[normalem]{ulem}
\usepackage{listings}
\usepackage[frozencache,cachedir={_minted-paper}]{minted}
\usepackage[inline]{enumitem}

\definecolor{heraldBlue}{rgb}{0.0,0.0,0.8}
\definecolor{heraldRed}{rgb}{0.8,0.0,0.0}
\definecolor{heraldGray}{rgb}{0.2,0.2,0.5}
\definecolor{heraldGreen}{rgb}{0.0,0.4,0.0}
\definecolor{heraldPink}{rgb}{0.8,0.1,0.6}

\newcommand{\sys}{FlexTOE\xspace}
\newcommand{\libsys}{libTOE\xspace}

\newcommand{\shortsecref}[1]{\S{}\ref{#1}}

\AtBeginDocument{\providecommand\BibTeX{{\normalfont B\kern-0.5em{\scshape i\kern-0.25em b}\kern-0.8em\TeX}}}

\acmConference[NSDI '22]{19th USENIX Symposium on Networked
  Systems Design and Implementation}{April 4--6, 2022}{Renton, WA}

\begin{document}
\title{\huge \sys: Flexible TCP Offload with Fine-Grained
Parallelism}

\author{Rajath Shashidhara$^1$ \hspace{4ex} Tim Stamler$^2$
  \hspace{4ex} Antoine Kaufmann$^3$ \hspace{4ex} Simon Peter$^1$\\
\large $^1$University of Washington \hspace{3ex} $^2$UT Austin \hspace{3ex} $^3$MPI-SWS}

\begin{abstract}
\sys is a flexible, yet high-performance TCP offload engine (TOE) to
  SmartNICs. \sys eliminates almost all host data-path TCP processing
  and is fully customizable. \sys interoperates well with other TCP
  stacks, is robust under adverse network conditions, and supports
  POSIX sockets.

\sys focuses on data-path offload of established connections,
  avoiding complex control logic and packet buffering in the NIC.
  \sys leverages fine-grained parallelization of the TCP data-path and
  segment reordering for high performance on wimpy SmartNIC
  architectures, while remaining flexible via a modular design. We
  compare \sys on an Agilio-CX40 to host TCP stacks Linux and TAS, and
  to the Chelsio Terminator TOE. We find that Memcached scales up to
  38\% better on \sys versus TAS, while saving up to 81\% host CPU
  cycles versus Chelsio. \sys provides competitive performance for
  RPCs, even with wimpy SmartNICs. \sys cuts 99.99th-percentile RPC
  RTT by 3.2$\times$ and 50\% versus Chelsio and TAS,
  respectively. \sys's data-path parallelism generalizes across
  hardware architectures, improving single connection RPC throughput
  up to 2.4$\times$ on x86 and 4$\times$ on BlueField.
\sys supports C and XDP programs written in eBPF. It
  allows us to implement popular data center transport features, such
  as TCP tracing, packet filtering and capture, VLAN stripping, flow
  classification, firewalling, and connection splicing.

\end{abstract}

\maketitle
\hypersetup{pdfauthor={Rajath Shashidhara, Tim Stamler,
                       Antoine Kaufmann, Simon Peter},
            pdfcreator={Rajath Shashidhara, Tim Stamler,
                        Antoine Kaufmann, Simon Peter},
            pdfpublisher={USENIX Association}}
\pagestyle{empty}

\section{Introduction}
TCP remains the default protocol in many networks, even as its CPU
overhead is increasingly a burden to application
performance~\cite{arrakis,ix,mtcp}. A long line of improvements to
software TCP stack architecture has reduced overheads: Careful packet
steering improves cache-locality for
multi-cores~\cite{affinity_accept,fastsocket,mtcp}, kernel-bypass
enables safe direct NIC access from user-space~\cite{ix,arrakis},
application libraries avoid system calls for common socket
operations~\cite{mtcp}, and fast-paths drastically reduce TCP
processing overheads~\cite{tas}.  Yet, even with these optimizations,
communication-intensive applications spend up to 48\% of per-CPU
cycles in the TCP stack and NIC driver (\S\ref{sec:tcp_overhead}).

Offload promises further reduction of CPU overhead. While moving parts
of TCP processing, such as checksum and segmentation, into the NIC is
commonplace~\cite{about_nics}, full TCP offload engines
(TOEs)~\cite{toe,chelsio,chimney} have so far failed to find
widespread adoption. A primary reason is that fixed
offloads~\cite{linux_toe} limit protocol evolution after
deployment~\cite{dumb_toe,accelnet,snap}. Tonic~\cite{tonic} provides
building blocks for flexible transport protocol offload to
FPGA-SmartNICs, but FPGA development is still difficult and slow.

We present \sys, a high-performance, yet flexible offload of the
widely-used TCP protocol. \sys focuses on scenarios that are common in
data centers, where connections are long-lived and small transfers are
common~\cite{snap}. \sys offloads the TCP data-path to a network
processor (NPU) based SmartNIC, enabling full customization of
transport logic and flexibility to implement data-path features whose
requirements change frequently in data centers. Applications interface
directly but transparently with the \sys datapath through the
\emph{\libsys} library that implements POSIX sockets, while \sys offloads
all TCP data-path processing (\shortsecref{sec:tcp_overhead}).

TCP data-path offload to SmartNICs is challenging.
SmartNICs support only restrictive programming models with stringent
per-packet time budgets and are geared towards massive
parallelism with wimpy cores~\cite{ipipe}.
They often lack timers, as well as floating-point and other
computational support, such as division. Finally, offload has to mask
high-latency operations that cross PCIe. On the other hand, TCP
requires computationally intensive and stateful code paths to track
in-flight segments, for reassembly and retransmission, and to perform
congestion control~\cite{tonic}. For each connection, the TCP
data-path needs to provide low processing tail latency and high
throughput and is also extremely sensitive to reordering.

Resolving the gap between TCP's requirements and SmartNIC hardware
capabilities requires careful offload design to efficiently utilize
SmartNIC capabilities. Targeting \sys at the TCP data-path of
established connections avoids complex control logic in the
NIC. \sys's offloaded data-path is one-shot for each TCP
segment---segments are never buffered in the NIC. Instead, per-socket
buffers are kept in per-process host memory where \libsys interacts
with them directly. Connection management, retransmission, and
congestion control are part of a separate control-plane, which
executes in its own protection domain, either on control cores of the
SmartNIC or on the host. To provide scalability and flexibility, we
decompose the TCP data-path into fine-grained modules that keep
private state and communicate explicitly. Like
microservices~\cite{snap}, \sys modules leverage a data-parallel
execution model that maximizes SmartNIC resource use \textbf{and}
simplifies customization. We organize \sys modules into a
\emph{data-parallel computation pipeline}. We also \emph{reorder} segments on-the-fly to support parallel,
out-of-order processing of pipeline stages, while enforcing in-order
TCP segment delivery.
To our
knowledge, no prior work attempting full TCP data-path offload to NPU
SmartNICs exists.

\noindent
We make the following contributions:

\begin{compactitem}[\labelitemi]
\item We characterize the CPU overhead of TCP data-path processing for
  common data center applications (\S\ref{sec:tcp_overhead}). Our
  analysis shows that up to 48\% of per-CPU cycles are spent in TCP
  data-path processing, even with optimized TCP stacks.

\item We present \sys, a flexible, high-performance TCP offload
  engine (\S\ref{sec:design}).
\sys leverages data-path processing with fine-grained parallelism
  for performance, but remains flexible via a modular design. We show
  how to decompose TCP into a data-path and a control-plane, and the
  data-path into a data-parallel pipeline of processing modules to
  hide SmartNIC processing and data access latencies.

\item We implement \sys on the Netronome Agilio-CX40 NPU SmartNIC
  architecture, as well as x86 and Mellanox BlueField (\S\ref{sec:impl}).
Using \sys design principles, we are the first to demonstrate that
  NPU SmartNICs can support scalable, yet flexible TCP data-path offload. Our code is available at \url{https://tcp-acceleration-service.github.io/FlexTOE}.

\item We evaluate \sys on a range of workloads and compare to Linux,
  the high-performance TAS~\cite{tas} network stack, and a Chelsio
  Terminator TOE~\cite{chelsio} (\S\ref{sec:eval}). We find that the
  Memcached~\cite{memcached} key-value store scales throughput up to
  38\% better on \sys than using TAS, while saving up to 81\% host CPU
  cycles versus Chelsio. \sys cuts 99.99th-percentile RPC RTT by
  3.2$\times$ and 50\% versus Chelsio and TAS respectively, 27\% higher
  throughput than Chelsio for bidirectional long flows,and an order of magnitude higher throughput under 2\% packet loss
  than Chelsio. We extend the \sys data-path with debugging and
  auditing functionality to demonstrate flexibility. \sys maintains
  high performance when interoperating with other network
  stacks. \sys's data-path parallelism generalizes across platforms,
  improving single connection RPC throughput up to 2.4$\times$ on x86
  and 4$\times$ on BlueField.
\end{compactitem}
 \section{Background}\label{sec:background}

We motivate \sys by analyzing TCP host CPU processing overheads of
related approaches (\S\ref{sec:tcp_overhead}). We then place \sys in
context of this and further related work
(\S\ref{sec:related}). Finally, we survey the relevant \emph{on-path}
SmartNIC architecture (\S\ref{sec:nic_arch}).

\subsection{TCP Impact on Host CPU Performance}\label{sec:tcp_overhead}

We quantify the impact of different TCP processing approaches on host
CPU performance in terms of CPU overhead, execution efficiency, and
cache footprint, when processing common RPC-based workloads. We do so
by instrumenting a single-threaded Memcached~\cite{memcached} server
application
using hardware performance counters (cf.~\S\ref{eval:testbed} for
details of our testbed). We use the popular
memtier\_benchmark~\cite{memtier-benchmarks} to generate the client
load, consisting of 32\,B keys and values, using as many clients as
necessary to saturate the server, executing closed-loop KV
transactions on persistent connections. Table~\ref{tab:overheads}
shows a breakdown of our server-side results, for each Memcached
request-response pair, into NIC driver, TCP/IP stack, POSIX sockets,
Memcached application, and other factors.

\begin{table}
\footnotesize
\begin{tabular}{lrrrrrrrr}
\multirow{2}{*}{Module} & \multicolumn{2}{c}{\textbf{Linux}} & \multicolumn{2}{c}{\textbf{Chelsio}} & \multicolumn{2}{c}{\textbf{TAS}} & \multicolumn{2}{c}{\textbf{FlexTOE}} \\
                        & kc                & \%             & kc                 & \%              & kc               & \%            & kc                 & \%              \\ \toprule
NIC driver              & 0.71              & 6              & 1.28               & 14              & 0.18             & 5             & 0                  & 0               \\
TCP/IP stack            & 4.25              & 35             & 0.40               & 4               & 1.44             & 43            & 0                  & 0               \\
POSIX sockets           & 2.48              & 21             & 2.61               & 29              & 0.79             & 23            & 0.74               & 44              \\
Application             & 1.26              & 10             & 1.31               & 16              & 0.85             & 26            & 0.89               & 53              \\
Other                   & 3.42              & 28             & 3.28               & 37              & 0.09             & 3             & 0.04               & 3               \\ \midrule
Total                   & 12.13             & 100            & 8.89               & 100             & 3.34             & 100           & 1.67               & 100             \\ \midrule
Retiring                & 4.60              & 38             & 2.43               & 27              & 1.66             & 48            & 0.77               & 46              \\
Frontend bound          & 3.53              & 29             & 1.52               & 17              & 0.46             & 13            & 0.34               & 21              \\
Backend bound           & 3.40              & 28             & 4.68               & 53              & 1.24             & 36            & 0.46               & 27              \\
Bad speculation         & 0.55              & 5              & 0.26               & 3               & 0.13             & 4             & 0.09               & 6               \\ \midrule
Instructions (k)        & \multicolumn{2}{l}{16.18}          & \multicolumn{2}{l}{8.14}             & \multicolumn{2}{l}{6.26}         & \multicolumn{2}{l}{2.93}             \\
IPC                     & \multicolumn{2}{l}{1.33}           & \multicolumn{2}{l}{0.92}             & \multicolumn{2}{l}{1.85}         & \multicolumn{2}{l}{1.75}             \\
Icache (KB)             & \multicolumn{2}{l}{47.50}          & \multicolumn{2}{l}{73.43}            & \multicolumn{2}{l}{39.75}        & \multicolumn{2}{l}{19.00}            \\ \end{tabular}
\caption{Per-request CPU impact of TCP processing.\vspace{-15pt}}
\label{tab:overheads}
\end{table}

\paragraph{In-kernel.} Linux's TCP stack is versatile but bulky,
leading to a large cache footprint, inefficient execution, and high
CPU overhead.
Stateless offloads~\cite{about_nics}, such as segmentation and generic
receive offload~\cite{82599}, reduce overhead for large transfers, but
they have minimal impact on RPC workloads dominated by short flows. We
find that Linux executes 12.13\,kc per Memcached request on average,
with only 10\% spent in the application. Not only does Linux have a
high instruction and instruction cache (Icache) footprint, but
privilege mode switches, scattered global state, and coarse-grained
locking lead to 62\% of all cycles spent in instruction fetch stalls
(frontend bound), cache and TLB misses (backend bound), and branch
mispredictions (cf.~\cite{tas}). These inefficiences result in 1.33
instructions per cycle (IPC), leveraging only 33\% of our 4-way issue
CPU architecture. Linux is, in principle, easy to modify, but kernel
code development is complex and security sensitive. Hence, introducing
optimizations and new network functionality to the kernel is often
slow~\cite{snap,nsaas,netkernel}.

\paragraph{Kernel-bypass.} Kernel-bypass, such as in mTCP~\cite{mtcp}
and Arrakis~\cite{arrakis}, eliminates kernel overheads by entrusting
the TCP stack to the application, but it has security
implications~\cite{kernelinterposeflexible}. TAS~\cite{tas} and Snap~\cite{snap} instead execute a protected
user-mode TCP stack on dedicated cores, retaining security and
performance. By eliminating kernel calls, TAS spends only 800\,cycles
in the socket API---31\% of Linux's API overhead. TAS also reduces TCP
stack overhead to 34\% of Linux. TAS reduces Icache footprint, front
and back-end CPU stalls, improving IPC by 40\% versus Linux, and
reducing the total per-request CPU impact to 27\% of Linux. However,
kernel-bypass still has significant
overhead. Only 26\% of per-request cycles are spent in Memcached---the remainder
is spent in TAS (breakdown in \S\ref{sec:tas_breakdown}).

\paragraph{Inflexible TCP offload.}
TCP offload can eliminate host CPU overhead for TCP processing. Indeed,
TOEs~\cite{toe} that offload the TCP data-path to the NIC have existed
for a long time. Existing approaches, such as the Chelsio
Terminator~\cite{chelsio}, hard-wire the TCP offload. The resulting
inflexibility prevents data center operators from adapting the TOE to
their needs and leads to a slow upgrade path due to long hardware
development cycles. For example, the Chelsio Terminator line has been
slow to adapt to RPC-based data center workloads.

Chelsio's inflexbility shows in our analysis. Despite drastically
reducing the host TCP processing cycles to 10\% of Linux and 28\% of
TAS, Chelsio's TOE only modestly reduces the total per-request CPU
cycles of Memcached by 27\% versus Linux and inflates them by
2.6$\times$ versus TAS. Chelsio's design requires interaction through
the Linux kernel, leading to a similar execution profile despite
executing 50\% fewer host instructions per request. In addition,
Chelsio requires a sophisticated TOE NIC driver, with
complex buffer management and
synchronization. Chelsio's design is inefficient for RPC processing
and leaves only 16\% of the total per-request cycles to
Memcached---6\% more than Linux and 10\% fewer than TAS.

\paragraph{\sys} \sys eliminates all host TCP stack overheads. \sys's
instruction (and Icache) footprint is at least 2$\times$ lower than
the other stacks, leading to an execution profile similar to TAS,
where 46\% of all cycles are spent retiring instructions. In addition,
53\% of all cycles can be spent in Memcached---an improvement of
2$\times$ versus TAS, the next best solution. The remaining cycles are
spent in the POSIX sockets API, which cannot be eliminated with TCP
offload.

\sys is also flexible, allowing operators to modify the TOE at
will. For example, we have modified the TCP data-path many times,
implementing many features that require TOE modification, including
scalable socket API implementations~\cite{affinity_accept,fastsocket},
congestion control protocols~\cite{dctcp,timely}, scalable flow
scheduling~\cite{carousel}, scalable PCIe communication
protocols~\cite{nvme}, TCP tracing~\cite{bpftrace}, packet filtering
and capture (tcpdump and PCAP), VLAN stripping, programmable flow
classification (eBPF~\cite{bpf}), firewalling, and connection splicing
similar to AccelTCP~\cite{acceltcp}. All of these features are
desirable in data centers and are adapted frequently.

\subsection{Related Work}\label{sec:related}

Beyond the TCP implementations covered in \S\ref{sec:tcp_overhead}, we
cover here further related work in SmartNIC offload, parallel packet
processing, and API and network protocol specialization.

\paragraph{SmartNIC offload.}
On-path SmartNICs (\S\ref{sec:nic_arch}), based on network processor
units (NPUs) and FPGAs, provide a suitable substrate for flexible
offload. Arsenic~\cite{arsenic} is an early example of flexible packet
multiplexing on a SmartNIC.
Microsoft's Catapult~\cite{catapult} offloads network management,
while Dagger~\cite{dagger} offloads RPC processing to
FPGA-SmartNICs. Neither offloads a transport protocol, like
TCP. AccelTCP~\cite{acceltcp} offloads TCP connection management and
splicing~\cite{tcp_splice} to NPU-SmartNICs, but keeps the TCP
data-path on the host using mTCP~\cite{mtcp}. Tonic~\cite{tonic}
demonstrates in simulation that high-performance, flexible TCP
transmission offload might be possible, but it stops short of
implementing full TCP data-path offload (including receiver
processing) in a non-simulated environment.
LineFS~\cite{linefs} offloads a distributed file system to an off-path
SmartNIC, leveraging parallelization to hide execution latencies of
wimpy SmartNIC CPUs and data access across PCIe. Taking inspiration
from Tonic and LineFS, but also from actor, and microservice-based
approaches presented in iPipe~\cite{ipipe}, E3~\cite{e3}, and
Click~\cite{click,li2016clicknp}, \sys shows how to decompose the TCP
data-path into a fine-grained data-parallel pipeline to support full
and flexible offload to on-path NPU-SmartNICs.

\paragraph{Parallel packet processing} RouteBricks~\cite{routebricks}
parallelizes across cores and cluster nodes for high-performance
routing, achieving high line-rates but remaining flexible via software
programmability. Routing relies on read-mostly state and is simple
compared to TCP. \sys applies fine-grained parallelization to complex,
stateful code paths.

\paragraph{Specialized APIs and protocols.} Another approach to lower
CPU utilization is specialization.  R2P2~\cite{r2p2} is a UDP-based
protocol for remote procedure calls (RPCs) optimized for efficient and
parallel processing, both at the end-hosts and in the network.
eRPC~\cite{erpc} goes a step further and co-designs an RPC protocol
and API with a kernel-bypass network stack to minimize CPU overhead
per RPC.
RDMA~\cite{rdma} is a popular combination of a networking API,
protocol, and a (typically hardware) network
stack. iWARP~\cite{iwarp}, in particular, leverages a TCP stack
underneath RDMA, offloading both.
These approaches improve processing efficiency, but at the cost of
requiring application re-design, all-or-nothing deployments, and
operational issues at scale~\cite{rdma_at_scale}, often due to
inflexibility~\cite{dumb_toe,linux_toe}.  \sys instead offloads the
TCP protocol in a flexible manner by relying on SmartNICs. Upper-layer
protocols, such as iWARP, can also be implemented using \sys.

\subsection{On-path SmartNIC Architecture}\label{sec:nic_arch}

On-path SmartNICs\footnote{Mellanox BlueField~\cite{bluefield} and
  Broadcom Stingray~\cite{stingray} are off-path SmartNICs that are
  not optimized for packet processing~\cite{ipipe}.},
such as Marvell Octeon~\cite{octeon}, Pensando Capri~\cite{pensando,pensando_datasheet}, and Netronome
Agilio~\cite{agiliocx,agiliolx}, support massively parallel packet
processing with a large pool of flow processing cores (FPCs), but they
lack efficient support for sophisticated program control flow and
complex computation~\cite{ipipe}.

\begin{wrapfigure}[13]{r}{.5\columnwidth}\centering \includegraphics[width=.5\columnwidth]{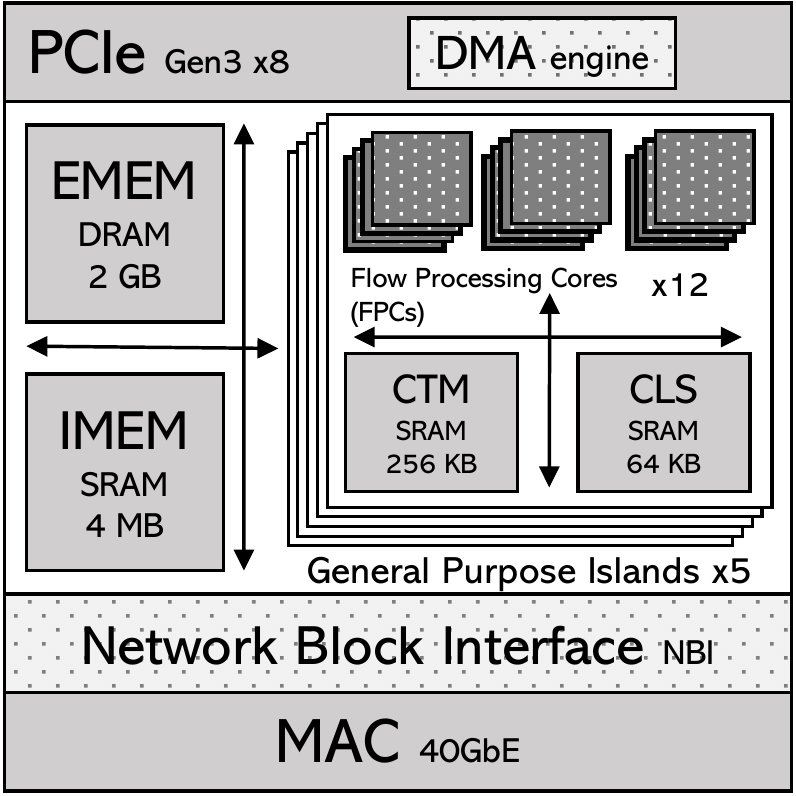}\caption{NFP-4000 overview.}\label{fig:nfp4000}\end{wrapfigure}

We explore offload to the NFP-4000 NPU, used in Netronome Agilio CX
SmartNICs~\cite{agiliocx}. We show the relevant architecture in
Figure~\ref{fig:nfp4000}. Like other on-path SmartNICs, FPCs are
organized into islands with local memory and processing
resources, akin to NUMA domains. Islands are connected in a mesh via a
high-band\-width interconnect (arrows in Figure~\ref{fig:nfp4000}).
The \emph{PCIe} island has up to two PCIe Gen3 x8 interfaces and a DMA
engine exposing DMA transaction queues ~\cite{PCIeUnderstanding}. FPCs can issue up to 256
asynchronous DMA transactions to perform IO between host and NIC memory.
The \emph{MAC}
island supports up to two 40\,Gbps Ethernet interfaces, accessed via a
\emph{network block interface} (NBI).

\paragraph{Flow Processing Cores (FPCs).} 60 FPCs are grouped into
five general-purpose islands (each containing 12 FPCs). Each FPC is an
independent 32-bit core at 800\,MHz with 8 hardware threads, 32\,KB
instruction memory, 4\,KB data memory, and CRC
acceleration. While FPCs have strong data flow processing capabilities, they have
small codestores, lack timers, as well as floating-point and other
complex computational support, such as division. This makes them
unsuitable to execute computationally and control intensive TCP
functionality, such as congestion, connection, and complex
retransmission control. For example, congestion avoidance involves
computing an ECN-ratio (gradient). We found that it takes 1,500 cycles
(1.9 $\mu$s) per RTT to perform this computation on FPCs.

\paragraph{Memory.} The NFP-4000 includes multiple memories of various
sizes and performance characteristics. General-purpose islands have
64KB of island-local scratch (\emph{CLS}) and 256\,KB of island target
memory (\emph{CTM}), with access latencies of up to 100 cycles from
island-local FPCs for data processing and transfer, respectively.  The
internal memory unit (\emph{IMEM}) provides 4\,MB of SRAM with an access
latency of up to 250 cycles. The external memory unit (\emph{EMEM})
provides 2\,GB of DRAM, fronted by a 3\,MB SRAM cache, with up to 500
cycles latency.

\paragraph{Implications for flexible offload.}
The NFP-4000 supports a broad range of protocols, but the computation
and memory restrictions require careful offload design.
As FPCs are wimpy and memory latencies high, sequential instruction
execution is much slower than on host processors.  Conventional
run-to-completion processing that assigns entire connections to
cores~\cite{tas,mtcp,ix} results in poor per-connection throughput and
latency. In some cases, it is beyond the feasible instruction and
memory footprint.  Instead, an efficient offload needs to leverage
more fine-grained parallelism to limit the per-core compute and memory
footprint.
 \section{\sys Design}\label{sec:design}

In addition to flexibility, \sys has the following goals:
\smallskip

\begin{compactitem}[\labelitemi]
\item \textbf{Low tail latency and high throughput}.  Modern
  datacenter network loads consist of short and long flows. Short
  flows, driven by remote procedure calls, require low tail completion
  time, while long flows benefit from high throughput. \sys shall
  provide both.

\item \textbf{Scalability}. The number of network flows and
  application contexts that servers must handle simultaneously is
  increasing. \sys shall scale with this demand.

\end{compactitem}

\vspace{1ex}
\noindent
To achieve these goals and overcome SmartNIC hardware limitations, we
propose three design principles:

\smallskip
\begin{compactenum}[1.]
\item \textbf{One-shot data-path offload.} We focus offload on the TCP
  RX/TX data-path, eliminating complex control, compute, and state,
  thereby also enabling fine-grained parallelization. Further,
  our data-path offload is one-shot for each TCP segment. Segments are
  never buffered on the NIC, vastly simplifying SmartNIC memory
  management.

\item \textbf{Modularity.} We decompose the TCP data-path into
  fine-grained, customizable modules that keep private state and
  communicate explicitly. New TCP extensions can be implemented as
  modules and hooked into the data-flow, simplifying development and
  integration.

\item \textbf{Fine-grained parallelism.} We organize the data-path
  modules into a data-parallel computation pipeline that maximizes
  SmartNIC resource use. We map stages to FPCs, allowing us to fully
  utilize all FPC resources. We employ TCP segment sequencing and
  reordering to support parallel, out-of-order processing of pipeline
  stages, while enforcing in-order segment delivery.

\end{compactenum}

\paragraph{Decomposing TCP for offload.}
We use the TAS host TCP stack architecture~\cite{tas}
as a starting point.
TAS splits TCP processing into three components: a data-path,
a control-plane, and an application library.
The data-path is responsible for scalable data transport of established
connections: TCP segmentation, loss detection and recovery, rate
control, payload transfer between socket buffers and the network, and
application notifications.
The control-plane handles connection and context management,
congestion control, and complex recovery involving timeouts.
Finally, the application library intercepts POSIX socket API
calls and interacts with control-plane and
data-path using dedicated context queues in shared memory.
Data-path and control-plane execute in their own protection domains on
dedicated cores, isolated from untrusted applications, and communicate
through efficient message passing queues.

\paragraph{\sys offload architecture.}
In \sys we adapt this architecture for offload, by designing and
integrating a \emph{data-path running efficiently} on the SmartNIC
(\S\ref{sec:tcp_parallel}).
The \sys control-plane can run on the host or on a SmartNIC control
CPU, with the same functionality as in TAS (cf.~\S\ref{sec:controlplane}).
The \sys control-plane additionally manages the SmartNIC data-path resources.
Similarly, our application library (\libsys) intercepts POSIX
socket calls and is dynamically linked to unmodified processes that
use \sys, and communicates directly with the data-path.

\begin{figure}
  \centering
\includegraphics[width=\columnwidth]{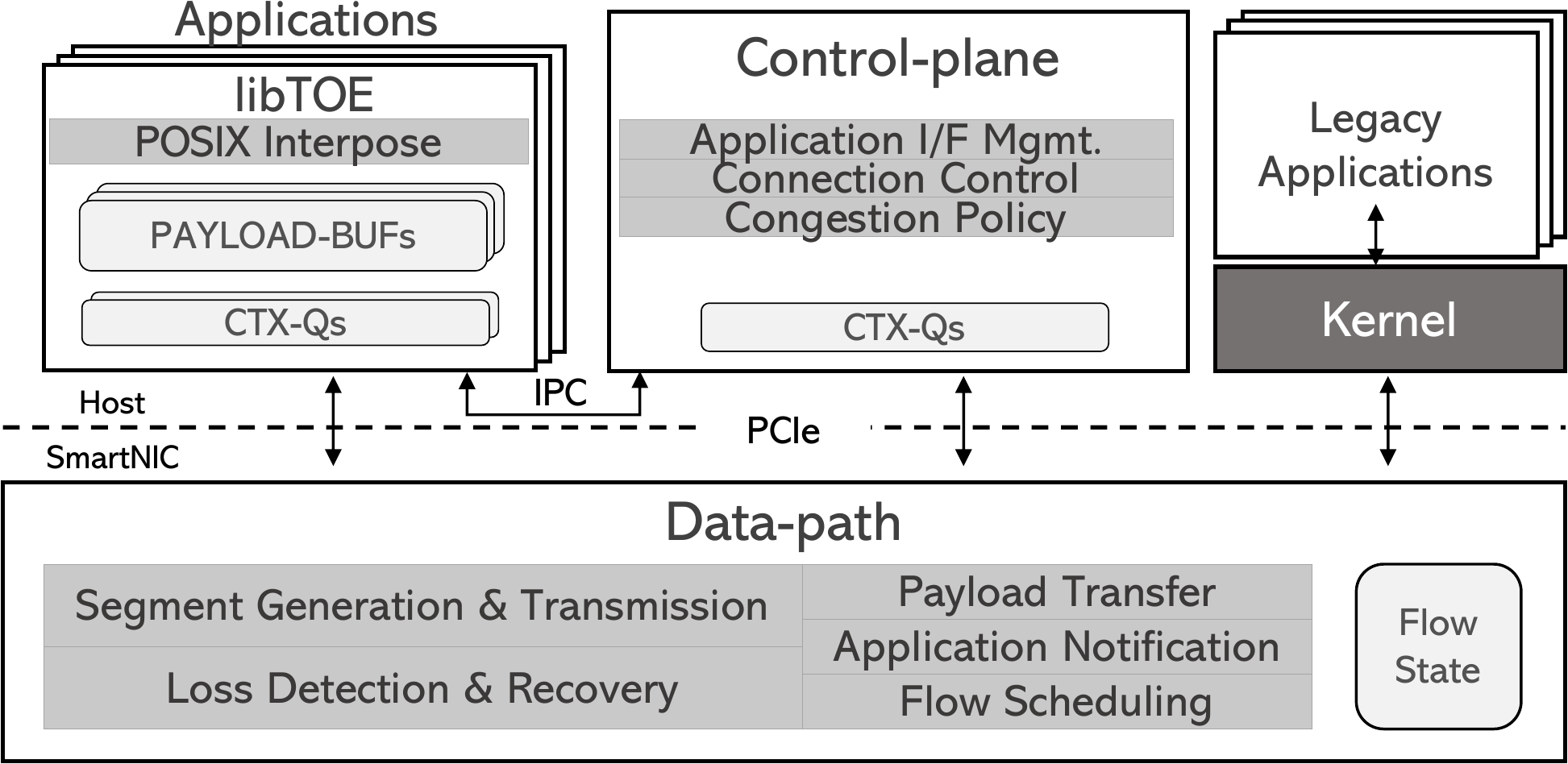}
  \caption{\sys offload architecture (host control-plane).}
  \label{fig:tasnic}
\end{figure}

Figure~\ref{fig:tasnic} shows the offload architecture of \sys, with a
host control-plane (each box is a protection domain).
\libsys, data-path, and control-plane communicate via pairs of
\emph{context queues} (CTX-Qs), one for each communication
direction.
CTX-Qs leverage PCIe DMA and MMIO or shared memory for SmartNIC-host
and intra-host communication, respectively.
\sys supports per-thread context queues for
scalability. Each TCP socket keeps receive and transmit payload buffers
(PAYLOAD-BUFs) in host memory. \libsys appends data for transmission
into the per-socket TX PAYLOAD-BUF and notifies the data-path using a
thread-local CTX-Q. The data-path appends received segments to the
socket's RX PAYLOAD-BUF after reassembly and \libsys is notified via
the same thread-local CTX-Q. Non-\sys traffic is forwarded to the Linux kernel, which legacy
applications may use simultaneously.

\subsection{TCP Data-path Parallelization}\label{sec:tcp_parallel}

To provide high offload performance using relatively wimpy SmartNIC
FPCs, \sys has to leverage all available parallelism within the TCP
data-path. In this section, we analyze the TAS host TCP
data-path to investigate what parallelism can be extracted. In
particular, the TCP data-path in TAS has the following three workflows:

\smallskip
\begin{compactitem}[\labelitemi]
\item{\textbf{Host control (HC)}}: When an application wants to
  transmit data, executes control operations on a socket, or when
  retransmission is necessary, the data-path must update the
  connection's transmit and receive windows accordingly.

\item{\textbf{Transmit (TX)}}: When a TCP connection is ready to
  send---based on congestion and flow control---the data-path prepares
  a segment for transmission, fetching its payload from a socket
  transmit buffer and sending it out to the MAC.

\item{\textbf{Receive (RX)}}: For each received segment of an
  established connection, the data-path must perform byte-stream
  re\-assembly---advance the TCP window, determine the segment's
  position in the socket receive buffer, generate an acknowledgment to
  the sender, and, finally, notify the application. If the received
  segment acknowledges previously transmitted segments, the data-path
  must also free the relevant payload in the socket transmit buffer.
\end{compactitem}
\smallskip

\noindent
Host TCP stacks, such as Linux or TAS, typically process each workflow
to completion in a critical section accessing a shared per-connection
state structure.
HC workflows are typically processed on the program threads that
trigger them, while TX and RX are typically triggered by NIC
interrupts and processed on high-priority (kernel or dedicated) threads.

\begin{figure}
    \centering
    \includegraphics[width=\columnwidth]{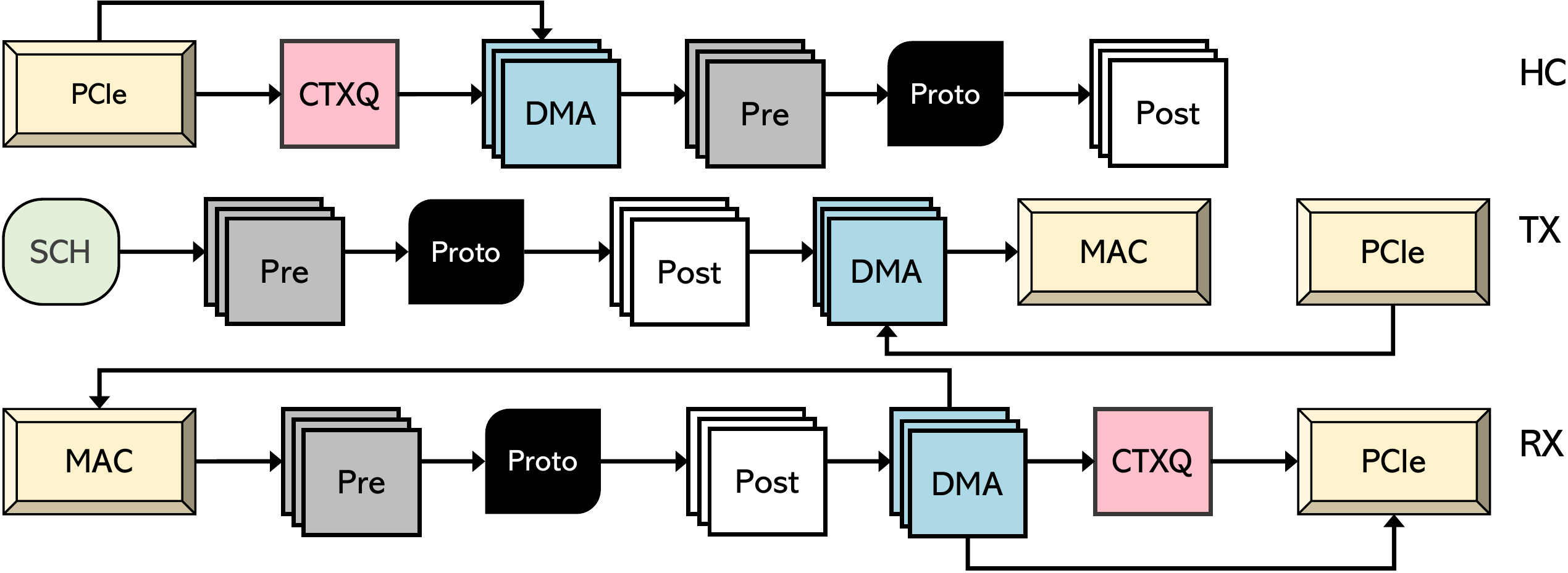}
    \caption{Per-connection data-path workflows. \emph{Protocol} is
      atomic. Other stages may be replicated for parallelism.}
    \label{fig:pipeline}
\end{figure}

For efficient offload, we decompose this data-path into an up to five-stage parallel
pipeline of processing modules: \emph{pre-processing},
\emph{protocol}, \emph{post-processing}, \emph{DMA}, and \emph{context
  queue} (Figure \ref{fig:pipeline}). Accordingly, we partition connection state into
module-local state (cf. \S\ref{app:state}).
The pipeline stages are chosen to maximize data-path
parallelism. Pre-processing accesses connection identifiers such as
MAC and IP addresses for segment header preparation and filtering. The
post-processing block handles application interface parameters, such
as socket buffer addresses and context queues. These parameters are
read-only after connection establishment and enable
coordi\-nation-free scaling. Congestion control statistics are
collected by the post-processor, but are only read by forward stages
and can be updated out-of-order (updates commute). The protocol stage
executes data-path code that must atomically modify protocol state,
such as sequence numbers and socket buffer positions. It is the only
\emph{pipeline hazard}---it cannot execute in parallel with other
stages. The DMA stage is stateless, while context queue stages may be
sharded. Both conduct high-latency PCIe transactions and are thus
separate stages that execute in parallel and scale independently.

We run pipeline stages on dedicated FPCs that utilize local memory for
their portion of the connection state. Pipelining allows us to execute
the data-path in parallel. It also allows us to replicate
processing-intensive pipeline stages to scale to additional FPCs. With
the exception of protocol processing, which is atomic per connection,
all pipeline stages are replicated. To concurrently process multiple
connections, we also replicate the entire pipeline. To keep flow state
local, each pipeline handles a fixed \emph{flow-group}, determined by
a hash on the flow's 4-tuple (the flow's protocol type is ignored---it
must be TCP). We now describe how we parallelize each data-path
workflow by decomposing it into these pipeline stages.

\subsubsection{Host Control (HC)}\label{sec:hcflow}

\begin{figure}
    \centering
    \includegraphics[width=\columnwidth]{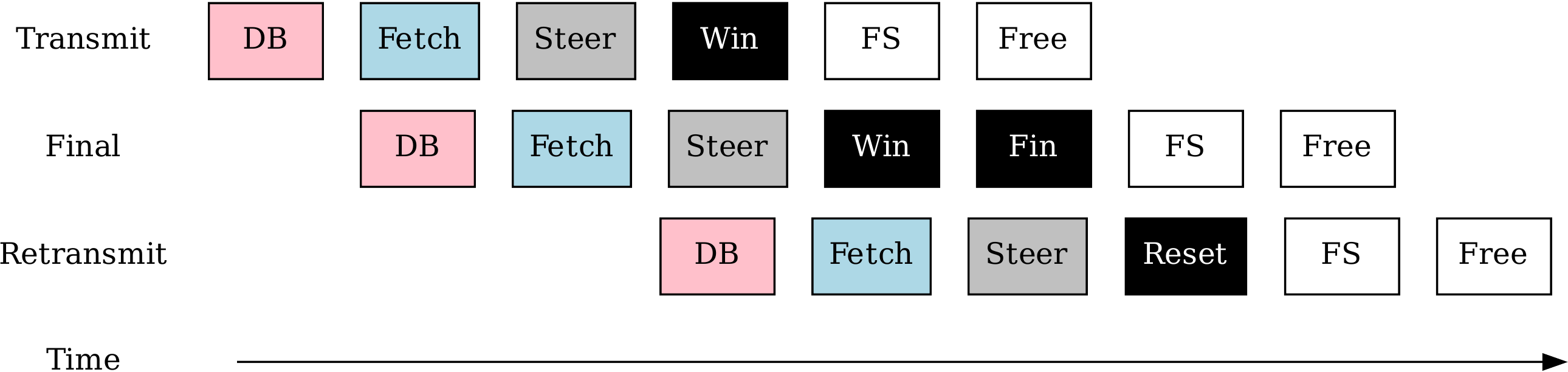}
    \caption{HC pipeline: Transmit, FIN, and retransmit.}
    \label{fig:aeflow}
\end{figure}

HC processing is triggered by a PCIe doorbell (DB) sent via
memory-mapped IO (MMIO) by the host to the context queue
stage. Figure~\ref{fig:aeflow} shows the HC pipeline for two transmits
(the second transmit closes the connection) triggered by \libsys, and
a retransmit triggered by the control-plane. HC requests may be
batched.

The context queue stage polls for DBs. In response to a DB, the stage
allocates a descriptor buffer from a pool in NIC memory. The limited
pool size flow-controls host interactions. If allocation fails,
processing stops and is retried later. Otherwise, the DMA stage
fetches the descriptor from the host context queue into the buffer
(Fetch). The pre-processor reads the descriptor, determines the
flow-group, and routes to the appropriate protocol stage (Steer). The
protocol stage updates connection receive and transmit windows
(Win). If the HC descriptor contains a connection-close indication,
the protocol stage also marks the connection as FIN (Fin). When the
transmit window expands due to the application sending data for
transmission, the post-processor updates the flow scheduler (FS) and
returns the descriptor to the pool (Free).

Retransmissions in response to timeouts are triggered by the
control-plane and processed the same as other HC events (fast
retransmits due to duplicate ACKs are described in
\S\ref{sec:rxflow}). The protocol stage resets the transmission state
(Reset) to the last ACKed sequence number (go-back-N retransmission).

\subsubsection{Transmit (TX)}\label{sec:tx_proc}

\begin{figure}
    \centering
    \includegraphics[width=\columnwidth]{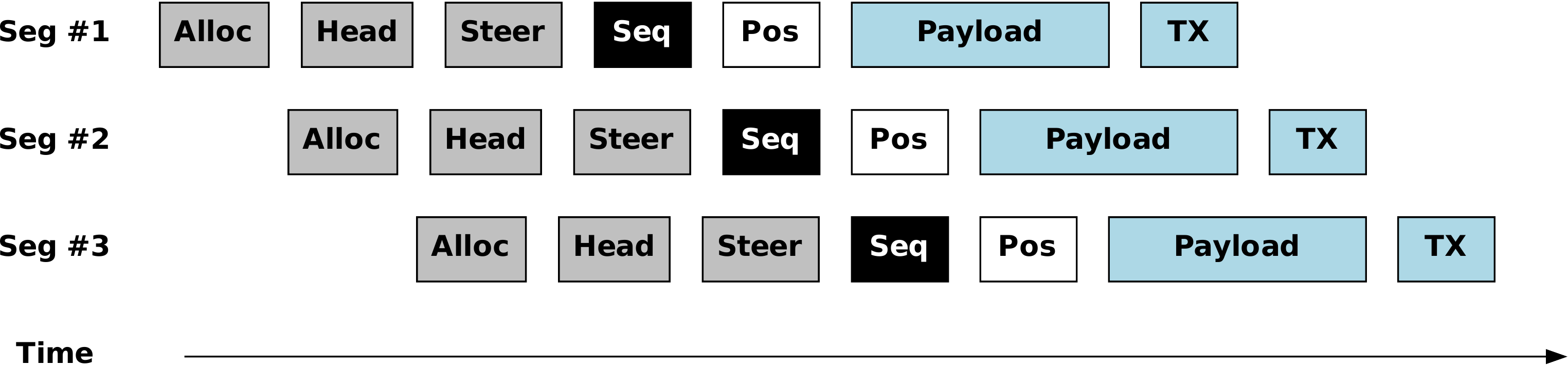}
    \caption{TX pipeline sending 3 segments.}
    \label{fig:txflow}
\end{figure}

Transmission is triggered by the flow scheduler (SCH) when a connection can
send segments. Figure~\ref{fig:txflow} shows the TX pipeline for 3
example segments.

The pre-processor allocates a segment in NIC memory (Alloc), prepares
Ethernet and IP headers (Head), and steers the segment to the
flow-group's protocol stage (Steer). The protocol stage assigns a TCP
sequence number based on connection state and determines the transmit
offset in the host socket transmit buffer (Seq). The post-processor
determines the socket transmit buffer address in host memory
(Pos). The DMA stage fetches the host payload into the segment
(Payload). After DMA completes, it issues the segment to the NBI (TX),
which transmits and frees it.

\subsubsection{Receive (RX)}\label{sec:rxflow}

\begin{figure}
    \centering
\includegraphics[width=\columnwidth]{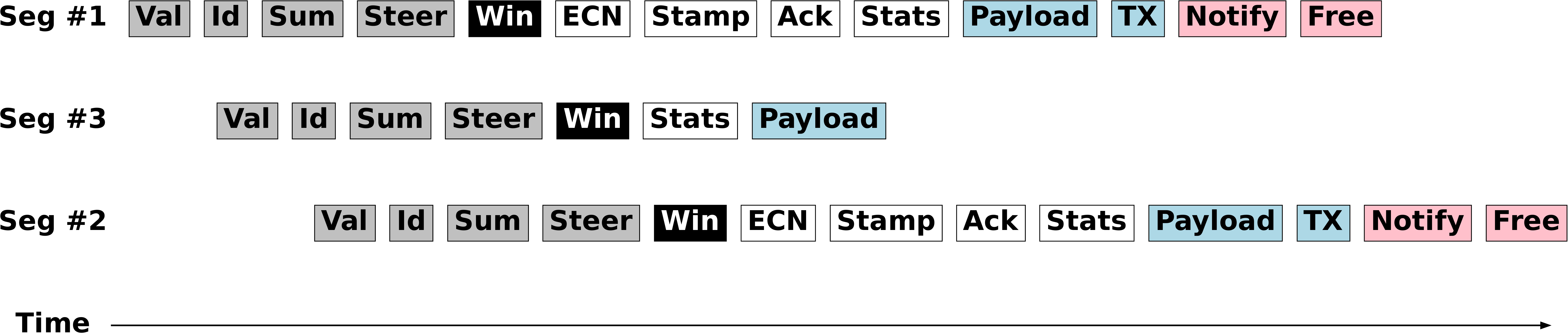}
    \caption{RX pipeline receiving 3 segments, 1 out of order.}
    \label{fig:rxflow}
\end{figure}

Figure~\ref{fig:rxflow} shows the RX pipeline for 3 example segments,
where segment \#3 arrives out of order.

\paragraph{Pre-processing} The pre-processor first validates the
segment header (Val). Non-data-path segments\footnote{\emph{Data-path
    segments} have any of the ACK, FIN, PSH, ECE, and CWR flags and
  they may have the timestamp option.} are filtered and forwarded to
the control-plane. Otherwise, the pre-processor determines the
connection index based on the segment's 4-tuple (Id) that is used by
later stages to access connection state.
The pre-processor generates a \emph{header summary} (Sum), including
only relevant header fields required by later pipeline
stages and steers the summary and connection identifier to the protocol stage
of its flow-group (Steer).

\paragraph{Protocol} Based on the header summary, the protocol stage
updates the connection's sequence and acknowledgment numbers, the
transmit window, and determines the segment's position in the host
socket receive payload buffer, trimming the payload to fit the receive
window if necessary (Win). The protocol stage also tracks duplicate ACKs and triggers fast retransmissions if necessary, by
resetting the transmission state to the last acknowledged position.
Finally, it forwards a snapshot of relevant connection
state to post-processing.

Out-of-order arrivals (segment \#3 in Figure~\ref{fig:rxflow}) need
special treatment. Like TAS~\cite{tas}, we track one out-of-order
interval in the receive window, allowing the protocol stage to perform
reassembly directly within the host socket receive buffer. We merge
out-of-order segments within the interval in the host receive
buffer. Segments outside of the interval are dropped and generate
acknowledgments with the expected sequence number to trigger
retransmissions at the sender. This design performs well under loss (cf. \S\ref{eval:loss}).

\paragraph{Post-processing} The post-processor prepares an
acknowledgment segment (Ack). \sys provides explicit congestion
notification (ECN) feedback and accurate timestamps for RTT estimation
(Stamp) in acknowledgments. It also collects congestion control and
transmit window statistics, which it sends to the control-plane and
flow scheduler (Stats). Finally, it determines the physical address of
the host socket receive buffer, payload offset, and length for the DMA
stage. If \libsys is to be notified, the post-processor allocates a
context queue descriptor with the appropriate notification.

\paragraph{DMA} The DMA stage first enqueues payload DMA descriptors
to the PCIe block (Payload). After payload DMA completes, the DMA
stage forwards the notification descriptor to the context queue
stage. Simultaneously, it sends the prepared acknowledgment segment to
the NBI (TX), which frees it after transmission. This ordering is
necessary to prevent the host and the peer from receiving
notifications before the data transfer to the host socket receive
buffer is complete.

\paragraph{Context queue}
If necessary, the context queue stage allocates an entry on the context queue and issues
the context queue descriptor DMA to notify \libsys of new
payload (Notify) and frees the internal descriptor buffer (Free).

\subsection{Sequencing and Reordering}\label{sec:reordering}

\begin{figure}
    \centering
    \includegraphics[width=\columnwidth]{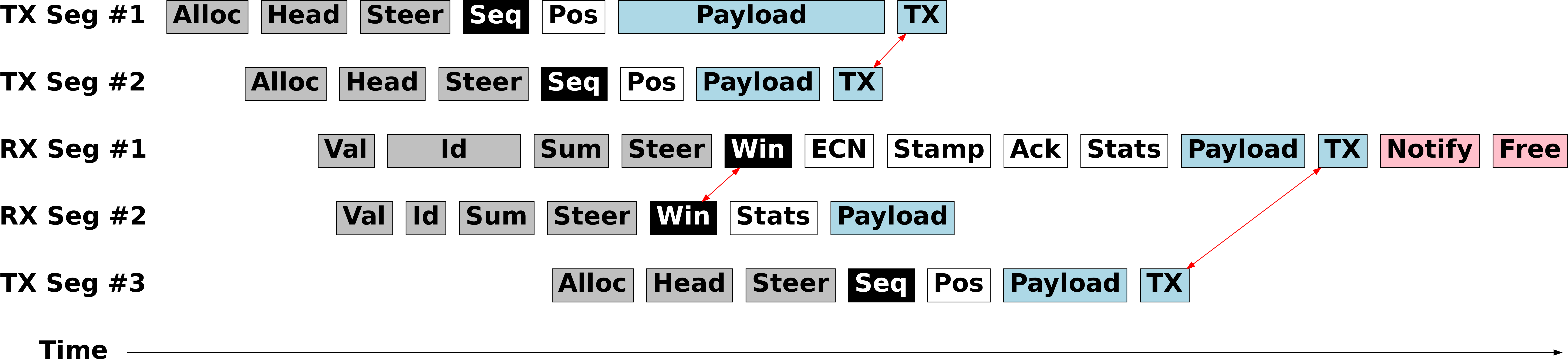}
    \caption{Undesirable pipeline reordering (red arrows).}
    \label{fig:tx_reorder}
\end{figure}

TCP requires that segments of the same connection are processed
in-order for receiver loss detection. However, stages in \sys's
data-parallel processing pipeline can have varying processing time and
hence may reorder segments. Figure~\ref{fig:tx_reorder} shows three
examples on a bidirectional connection where undesirable segment
reordering occurs.

\smallskip
\begin{compactenum}[1.]
\item \textbf{TX.} TX segment \#1 stalls in DMA across a congested
  PCIe link, causing it to be transmitted on the network after TX
  segment \#2, potentially triggering receiver loss detection.

\item \textbf{RX.} RX segment \#1 stalls in flow identification during
  pre-processing, entering the protocol stage later than RX segment
  \#2. The protocol stage detects a hole and triggers unnecessary
  out-of-order processing.

\item \textbf{ACK.} TX segment \#3 is processed after RX segment \#1
  in the protocol stage. RX segment \#1 generates an ACK, but RX
  post-processing is complex, resulting in TX segment \#3 with a
  higher sequence number being sent before ACK segment \#1.
\end{compactenum}
\smallskip

\noindent
To avoid reordering, \sys's data-path pipeline sequences and reorders
segments if necessary. In particular, we assign a sequence number to
each segment entering the pipeline. The parallel pipeline stages can
operate on each segment in any order. The protocol stage requires
in-order processing and we buffer and re-order segments that arrive
out-of-order before admitting them to the protocol stage. Similarly,
we buffer and re-order segments for transmission before admitting them
to the NBI. We leverage additional FPCs for sequencing, buffering, and
reordering.

\subsection{Flexibility}\label{sec:flexibility}

Data center networks evolve quickly, requiring TCP stacks to be easily
modifiable by operators, not just
vendors~\cite{snap,nsaas,netkernel}. Many desirable data center
features require TOE modification and are adapted frequently by
operators. \sys provides flexibility necessary to implement and
maintain these features even beyond host stacks such as TAS, by
relying on a programmable SmartNIC. To simplify development and modification of the TCP data-path, \sys
provides an extensible, data-parallel pipeline of self-contained
modules, similar to the Click~\cite{click} extensible router.

\paragraph{Module API} The \sys module API provides developers
one-shot access to TCP segments and associated meta-data. Meta-data
may be created and forwarded along the pipeline by any module. Modules
may also keep private state. For scalability, private state cannot be
accessed by other modules or replicas of the same module. Instead,
state that may be accessed by further pipeline stages is forwarded as
meta-data.

The replication factor of pipeline stages and assignment to FPCs is
manual and static in \sys. As long as enough FPCs are available, this
approach is acceptable. Operators can determine an appropriate
replication factor that yields acceptable TCP processing bandwidth for
a pipeline stage via throughput microbenchmarks at deployment. Stages
that modify connection state atomically may be deployed by inserting
an appropriate steering stage that steers segments of a connection to
the module in the atomic stage, holding their state (cf. protocol
processing stage in \S\ref{sec:tcp_parallel}).

\paragraph{XDP modules} \sys also supports eXpress Data Path (XDP)
modules~\cite{xdpiovisor, netronome-ebpf-1, netronome-ebpf-2},
implemented in eBPF. XDP modules operate on raw packets, modify them if necessary, and
output one of the following result codes:
\begin{enumerate*}[label=(\roman*)]
\item \texttt{XDP\_PASS}: Forward the packet to the next \sys pipeline
  stage.
\item \texttt{XDP\_DROP}: Drop the packet.
\item \texttt{XDP\_TX}: Send the packet out the MAC.
\item \texttt{XDP\_REDIRECT}: Redirect the packet to the control-plane.
\end{enumerate*}

XDP modules may use BPF maps (arrays, hash tables) to store and modify
state atomically~\cite{bpflinuxman}, which may be modified by the
control-plane. For example, a firewall module may store blacklisted
IPs in a hash map and the control-plane may add or remove entries
dynamically. The module can consult the hash map to determine if a
packet is blacklisted and drop it. XDP stages scale like other
pipeline stages, by replicating the module. \sys automatically
reorders processed segments after a parallel XDP stage
(\S\ref{sec:reordering}).

Using these APIs, we modified the \sys data-path many times,
implementing the features listed in \S\ref{sec:tcp_overhead}
(evaluation in \S\ref{eval:flexibility}). Further, ECN feedback and
segment timestamping (cf.~\S\ref{sec:rxflow}) are optional TCP
features that support our congestion control policies. Operators can
remove the associated post-processing modules if they are not
needed.

By handling atomicity, parallelization, and ordering concerns, \sys
allows complex offloads to be expressed using few lines of code. For
example, we implement AccelTCP's connection splicing in 24 lines of eBPF code
(cf.~Listing~\ref{lst:splicexdp} in the appendix). The module performs a lookup on the
segment 4-tuple in a BPF hashmap. If a match is not found, we forward
the segment to the next pipeline stage. Otherwise, we modify the
destination MAC and IP addresses, TCP ports, and translate sequence
and acknowledgment numbers using offsets configured by the
control-plane, based on the connection's initial sequence
number. Finally, we transmit. \sys handles sequencing and updating the
checksum of the segment. Additionally, when we receive segments with
control flags indicating connection closure, we atomically remove the
hashmap entry and notify the control-plane.

\subsection{Flow Scheduling}\label{sec:queueman}

\sys leverages a work-conserving flow scheduler on the NIC
data-path. The flow scheduler obeys transmission rate-limits and
windows configured by the control-plane's congestion control
policy. For each connection, the flow scheduler keeps track of how
much data is available for transmission and the configured
rate. Transmission rates and windows are stored in NIC memory and are
directly updated by the control-plane using
MMIO.

We implement our flow scheduler based on
Carousel~\cite{carousel}. Carousel schedules a large number of flows
using a time wheel. Based on the next transmission time, as computed
from rate limits and windows, we enqueue flows into corresponding
slots in the time wheel. As the time slot deadline passes, the flow
scheduler schedules each flow in the slot for transmission
(\S\ref{sec:tx_proc}). To conserve work, the flow scheduler only adds
flows with a non-zero transmit window into the time wheel and bypasses
the rate limiter for uncongested flows. These flows are scheduled
round-robin.
 \section{Agilio-CX40 Implementation}\label{sec:impl}
This section describes \sys's Agilio-CX40 implementation. Due to space
constraints, the x86 and BlueField ports are described in detail in
\S\ref{sec:ports}. \sys's design across the different ports is
identical. We do not merge or split any of the fine-grained modules or
reorganize the pipeline across ports.

\sys is implemented in 18,008 lines of C code (LoC). The offloaded
data-path comprises 5,801 lines of C code. We implement parts of the
data-path in assembly for performance. \libsys contains 4,620 lines of
C, whereas the control path contains 5,549 lines of C. \libsys and the
control plane are adapted from TAS.  We use the NFP compiler toolchain
version 6.1.0.1 for SmartNIC development.

\paragraph{Driver} We develop a Linux \sys driver based on the
\texttt{igb\_uio} driver that enables \libsys and the control plane to
perform MMIO to the SmartNIC from user space. The driver supports
MSI-X based interrupts. The control-plane registers an
\texttt{eventfd} for each application context in the driver. The
interrupt handler in the driver pings the corresponding
\texttt{eventfd} when an interrupt is received from the data-path for
the application context. This enables \libsys to sleep when waiting
for IO and reduces the host CPU overhead of polling.

\paragraph{Host memory mapping} To simplify virtual to physical address translation for DMA
operations, we allocate physically contiguous host memory using 1G
hugepages. The control-plane maps a pool of 1G hugepages at startup
and allocates socket buffers and context queues out of this pool. In
the future, we can use the IOMMU to eliminate the requirement of
physically contiguous memory for \sys buffers.

\paragraph{Context queues} Context queues use shared memory on the
host, but communication between SmartNIC and host requires PCIe. We
use scalable and efficient PCIe communication techniques \cite{nvme}
that poll on host memory locations when executing in the host and on
NIC-internal memory when executing on the NIC. The NIC is notified of
new queue entries via MMIO to a NIC doorbell.
The context queue manager notifies applications through MSI-X
interrupts, converted by the driver to an \texttt{eventfd}, after a
queue has been inactive.

\subsection{Near-memory Processing}\label{sec:near_memory}

An order of magnitude difference exists in the access latencies of
different memory levels of the NFP-4000. For performance, it is
critical to maximize access to local memory. The NFP-4000 also provides certain near-memory acceleration,
including a lookup engine exposing a content addressable memory
(CAM) and a hash table for
fast matching, a queue memory engine exposing concurrent data structures such
as linked lists, ring buffers, journals, and work-stealing
queues. Finally, synchronization primitives such as ticket
locks and inter-FPC signaling are exposed to coordinate threads and to sequence packets. We build specialized caches at multiple levels in the
different pipeline stages using these primitives. Other NICs have
similar accelerators.

\paragraph{Caching} We use each FPC's CAM to build 16-entry
fully-associative local memory caches that evict entries based on
LRU. The protocol stage adds a 512-entry direct-mapped second-level
cache in CLS. Across four islands, we can accommodate up to 2K flows
in this cache. The final level of memory is in EMEM.  When an FPC processes a segment, it fetches the relevant state into
its local memory either from CLS or from EMEM, evicting other cache
entries as necessary.
We allocate connection identifiers in such a way that we minimize
collisions on the direct-mapped CLS cache.

\paragraph{Active connection database} To facilitate connection index
lookup in the pre-processing stage, we employ the hardware lookup
capability of IMEM to maintain a database of active connections. CAM
is used to resolve hash collisions. The pre-processor computes a
CRC-32 hash on a segment's 4-tuple to locate the connection index
using the lookup engine. The pre-processor caches up to 128 lookup
entries in its local memory via a direct-mapped cache on the hash
value.

\paragraph{FPC mapping} \sys's pipeline fully leverages the Agilio CX40 and is extensible to
further FPCs, e.g. of the Agilio LX~\cite{agiliolx}. For island-local
interactions among modules, we use CLS ring buffers. CLS supports the
fastest intra-island producer-consumer mechanisms. Among islands, we
rely on work-queues in IMEM and EMEM.

We use all but one general-purpose islands for the first three stages
of the data-path pipeline (\emph{protocol islands}). Each
island manages a \emph{flow-group}. While protocol and
post-processing FPCs are local to a flow-group, pre-processors
handle segments for any flow. We assign 4 FPCs to pre-/post-processing stages in each flow-group. Each island retains 3
unassigned FPCs that can run additional data-path modules
(\S\ref{eval:flexibility}).

On the remaining general-purpose island (called \emph{service
  island}), we host remaining pipeline stages and adjacent modules,
such as context queue FPCs, the flow scheduler (SCH), and DMA
managers. DMA managers are replicated to hide PCIe latencies. The
number of FPCs assigned to each functionality is determined such that
no functionality may become a bottleneck. Sequencing and reordering
FPCs are located on a further island with miscellaneous functionality.

\paragraph{Flow scheduler.} We implement Carousel using hardware
queues in EMEM. Each slot is allocated a hardware queue. To add a flow
to the time wheel, we enqueue it on the queue associated with
the time slot. Note that the order of flows within a particular slot
is not preserved. EMEM support for a large number of hardware queues
enables us to efficiently implement a time wheel with a small slot
granularity and large horizon to achieve high-fidelity congestion
control. Converting transmission rates to deadlines requires division,
which is not supported on the NFP-4000. Thus, the
control-plane computes transmission intervals in cycles/byte units
from rates and programs them to NIC memory. This enables the flow
scheduler to compute the time slot using only multiplication.
 \section{Evaluation}\label{sec:eval}

We answer the following evaluation questions:

\smallskip
\begin{compactitem}[\labelitemi]
\item \textbf{Flexible offload.} Can flexible offload improve
  throughput, latency, and scalability of data center applications?
Can we implement common data center features? (\S\ref{eval:kv})

\item \textbf{RPCs.} How does \sys's data-path parallelism enable TCP
  offload for demanding RPCs? Do these benefits generalize across
  hardware architectures? Does \sys provide low latency for short
  RPCs? Does \sys provide high throughput for long RPCs? To how many
  simultaneous connections can \sys scale?  (\S\ref{eval:rpc})

\item \textbf{Robustness.} How does \sys perform under loss and
  congestion? Does it provide connection-fairness? (\S\ref{eval:loss})

\end{compactitem}

\paragraph{Testbed cluster} \label{eval:testbed} Our evaluation setup
consists of two 20-core Intel Xeon Gold 6138 @ 2\,GHz machines, with 40\,GB RAM and 48\,MB aggregate cache. Both machines are equipped with
Netronome Agilio CX40 40\,Gbps (single port), Chelsio Terminator
T62100-LP-CR 100\,Gbps and Intel XL710 40\,Gbps NICs. We use one of
the machines as a server, the other as a client. As additional
clients, we also use two 2$\times$18-core Intel Xeon Gold 6154 @ 3\,GHz
systems with 90\,MB aggregate cache and two 4-core Intel Xeon E3-1230
v5 @ 3.4\,GHz systems with 9\,MB aggregate cache. The Xeon Gold machines
are equipped with Mellanox ConnectX-5 MT27800 100\,Gbps NICs, whereas
the Xeon E3 machines have 82599ES 10\,Gbps NICs. The machines are
connected to a 100\,Gbps Ethernet switch.

\paragraph{Baseline} We compare \sys performance against the
Linux TCP stack, Chelsio's kernel-based TOE\footnote{Chelsio does not
  support kernel-bypass.}, and the TAS kernel-bypass
stack\footnote{TAS~\cite{tas} performs better than mTCP~\cite{mtcp} on
  all of our benchmarks. Hence, we omit a comparison to mTCP and
  AccelTCP~\cite{acceltcp}, which uses mTCP.}. TAS does not perform
well with the Agilio CX40 due to a slow NIC DPDK driver. We run TAS on
the Intel XL710 NIC, as in \cite{tas}, unless mentioned otherwise. We
use identical application binaries across all baselines. DCTCP is our
default congestion control policy.

\subsection{Benefit of Flexible Offload}\label{eval:kv}\label{eval:offload}\label{eval:flexibility}

\begin{figure}
  \centering
  \includegraphics[width=\columnwidth]{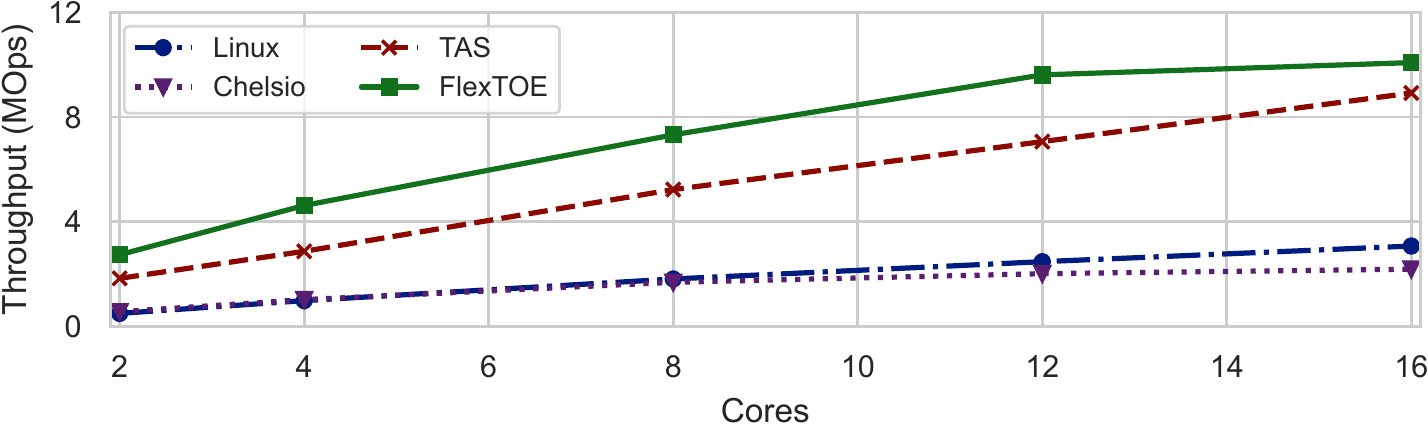}
  \caption{Memcached throughput scalability.}
\label{fig:core-scalability}
\end{figure}

\paragraph{Application throughput scalability.} Offloaded CPU cycles
may be used for application work. We quantify these benefits by
running a Memcached server, as in \S\ref{sec:tcp_overhead}, varying
the number of server cores. Figure~\ref{fig:core-scalability} shows that, by saving host CPU
cycles (cf.~Table~\ref{tab:overheads}), \sys achieves up to
1.6$\times$ TAS, 4.9$\times$ Chelsio, and 5.5$\times$ Linux throughput.
\sys and TAS scale similarly---both use per-core context
queues. The Agilio CX becomes a compute-bottleneck at 12 host cores. Linux and Chelsio are slow for this workload, due to system call
overheads, and do not scale well due to in-kernel locks.

\begin{figure}
  \centering
\includegraphics[width=\columnwidth]{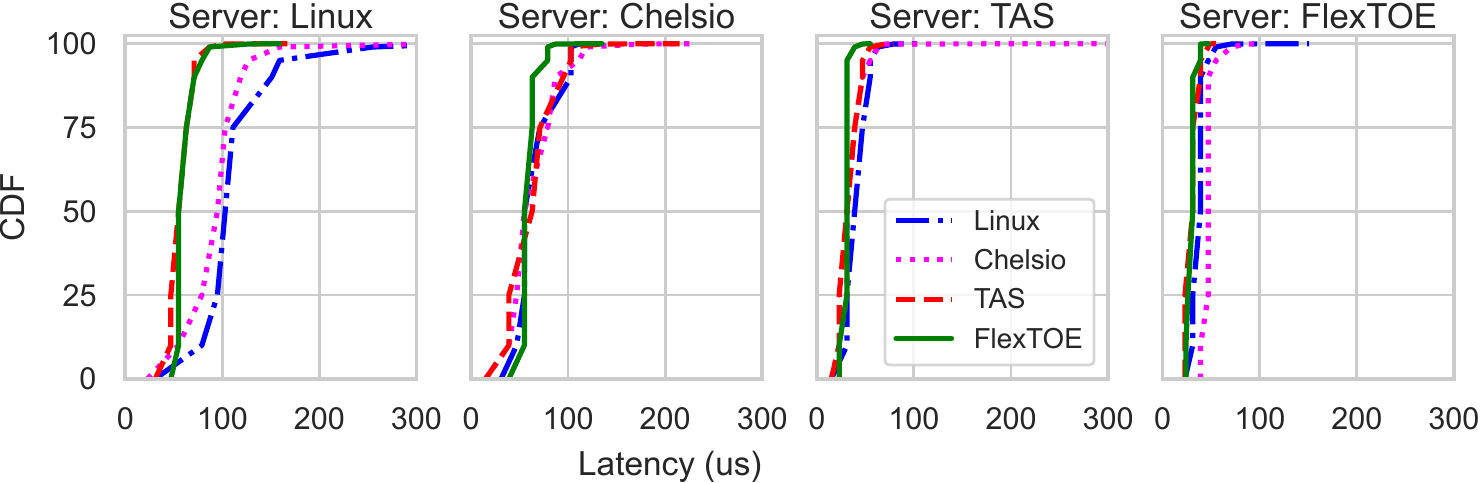}
  \caption{Latency of different server-client
    combinations.}
\label{fig:kv-latency}
\end{figure}

\paragraph{Low (tail) latency.} We repeat a single-threaded version of
the same Memcached benchmark for all server-client network stack
combinations. Latency distributions are shown in
Figure~\ref{fig:kv-latency}. We can see that \sys consistently
provides the lowest median and tail Memcached operation latency across
all stack combinations. Offload provides excellent performance
isolation by physically separating the TCP data-path, even though
\sys's pipelining increases minimum latency in some cases
(cf.~\S\ref{eval:rpc}).

\paragraph{Flexibility}
Unlike fixed offloads and in-kernel stacks, \sys provides full
user-space programmability via a module API, simplifying
development. Customizing \sys is simple and does not require a system
reboot. For example, we have developed logging, statistics, and profiling
capabilities that can be turned on only when necessary.
We make use of these capabilities during development and optimization
of \sys.
We implemented up to 48 different tracepoints (including examples from
bpftrace~\cite{bpftrace}) in the data-path pipeline, tracking
transport events such as per-connection drops, out-of-order packets
and retransmissions, inter-module queue occupancies, and critical
section lengths in the protocol module for various event
types. Table~\ref{tab:perfdebug} shows that profiling degrades data-path
performance versus the baseline by up to 24\% when all 48 tracepoints
are enabled. We also implement \texttt{tcpdump}-style traffic logging, including
packet filters based on header fields. Logging naturally has high
overhead (up to 43\% when logging all packets). \sys provides the
flexibility to implement these features and to turn them on only when
necessary.

Furthermore, new data-plane functionality leveraging the XDP API may
be dynamically loaded into \sys as eBPF programs. eBPF programs can be
compiled to NFP assembly. This level of dynamic flexibility is hard to
achieve with an FPGA as it requires instruction set programmability
(overlays~\cite{kernelinterposeflexible}). We measure the overhead of \sys XDP support by running a null program
that simply passes on every packet without modification. We observe
only 4\% decline in throughput. Common XDP modules, such as stripping
VLAN tags on ingress packets, also have negligible overhead. Finally, connection splicing (cf.~Listing~\ref{lst:splicexdp} in
the appendix) achieves a maximum splicing performance of 6.4 million
packets per second, enough to saturate the NIC line rate with
MTU-sized packets, leveraging only idle FPCs\footnote{We are
  compute-limited by our Agilio CX. Using an Agilio LX, like AccelTCP,
  would allow us to achieve even higher throughput.}.

\begin{table}[t]
\small
\begin{tabular}{lr}
\textbf{Build} & \textbf{Throughput} (MOps)\\
\toprule
Baseline \sys         & 11.35 \\
Statistics and profiling & 8.67 \\
\texttt{tcpdump} (no filter) & 6.52 \\
\texttt{XDP} (null) & 10.87 \\
\texttt{XDP} (vlan-strip) & 10.83 \\
\end{tabular}
\caption{Performance with flexible extensions.\vspace{-12pt}}
\label{tab:perfdebug}
\end{table}

\subsection{Remote Procedure Calls (RPCs)}
\label{eval:rpc}

RPCs are an important but difficult workload for flexible offload.
Latency and client scalability requirements favor fast processing
engines with large caches, such as found in CPUs and ASICs. Neither
are available in on-path SmartNICs. We show that flexible offload can
be competitive with state-of-the-art designs. We then show that \sys's
data-path parallelism is necessary to provide the necessary
performance.

\paragraph{Typical RX / TX performance.} We start with a typical server
scenario, processing RPCs of many (128) connections, produced in an
open loop by multiple (16) clients (multiple pipelined RPCs per
connection). To simulate application processing, our server waits for
an artificial delay of 250 or 1,000 cycles for each RPC. We run
single-threaded to avoid the network being a bottleneck. We quantify
RX and TX throughput separately, by switching RPC consumer and
producer roles among clients and servers, over different RPC sizes.

\begin{figure}
  \centering
\includegraphics[width=\columnwidth]{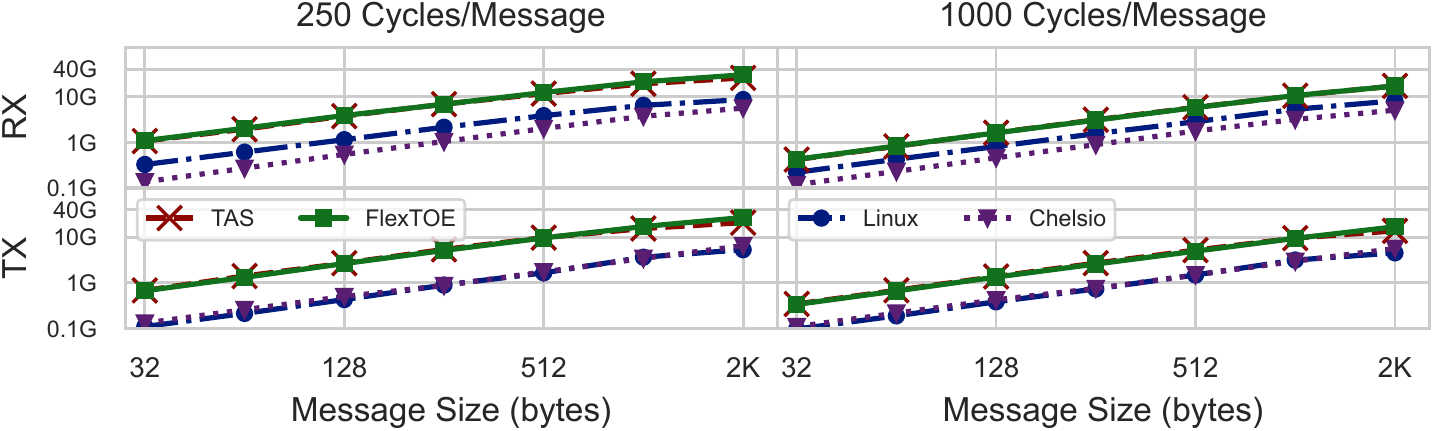}
  \caption{RPC throughput for saturated server.}
\label{fig:pipelined-rpc}
\end{figure}

Figure~\ref{fig:pipelined-rpc} shows the results. For 250 cycles of
processing overhead, \sys provides up to 4$\times$ better throughput
than Linux and 5.3$\times$ better throughput than Chelsio when
receiving. For 2\,KB message size, both TAS and \sys reach 40\,Gbps line
rate, whereas Linux and Chelsio barely reach 10\,Gbps and 7\,Gbps,
respectively. When sending packets, the difference in performance
between Linux and \sys is starker. \sys shows over 7.6$\times$ higher
throughput over both Linux and Chelsio for all message sizes. The
gains remain at over 2.2$\times$ as we go to 1,000
cycles/RPC. Performance of TAS and \sys track closely for all message
sizes. This is expected as the single application server core is
saturated by both network stacks (TAS runs on additional host cores).

\begin{figure}
  \centering
  \includegraphics[width=\columnwidth]{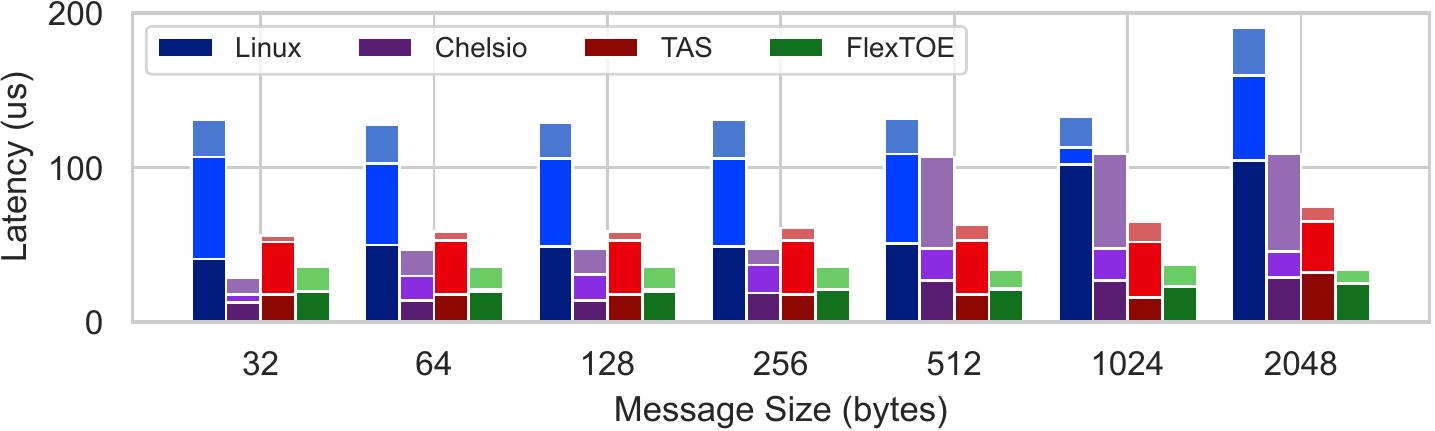}
  \caption{Median, 99p and 99.99p RPC RTT.}
\label{fig:short-rpc-latency}
\end{figure}

\medskip
\noindent
We break down this result by studying the performance sensitivity of
each TCP stack, varying each RPC parameter within its sensitive
dynamic range. For these benchmarks, we evaluate the raw performance
of the stacks, without application processing delays.

\paragraph{RPC latency} A client establishes a single connection to
the server and measures single RPC
RTT. Figure~\ref{fig:short-rpc-latency} shows the median and tail RTT
for various small message sizes (stacked bars). The inefficiency of
in-kernel networking is reflected in the median latency of Linux,
which is at least 5$\times$ worse compared to other stacks. For
message sizes < 256\,B, \sys's median latency (20\,us) is 1.4$\times$
Chelsio's median latency (14\,us) and 1.25$\times$ TAS's median latency
(16\,us). \sys's data-path pipeline across many wimpy FPCs increases
median latency for single RPCs. However, \sys has an up to 3.2$\times$
smaller tail compared to Chelsio and nearly constant per-segment
overhead as the RPC size increases. In case of a 2\,KB RPC (larger than
the TCP maximum segment size), \sys's latency distribution remains
nearly unchanged. \sys's fine-grain parallelism is able to hide the
processing overhead of multiple segments, providing 22\% lower median
and 50\% lower tail latency than TAS.

\begin{figure}
  \centering
  \includegraphics[width=\columnwidth]{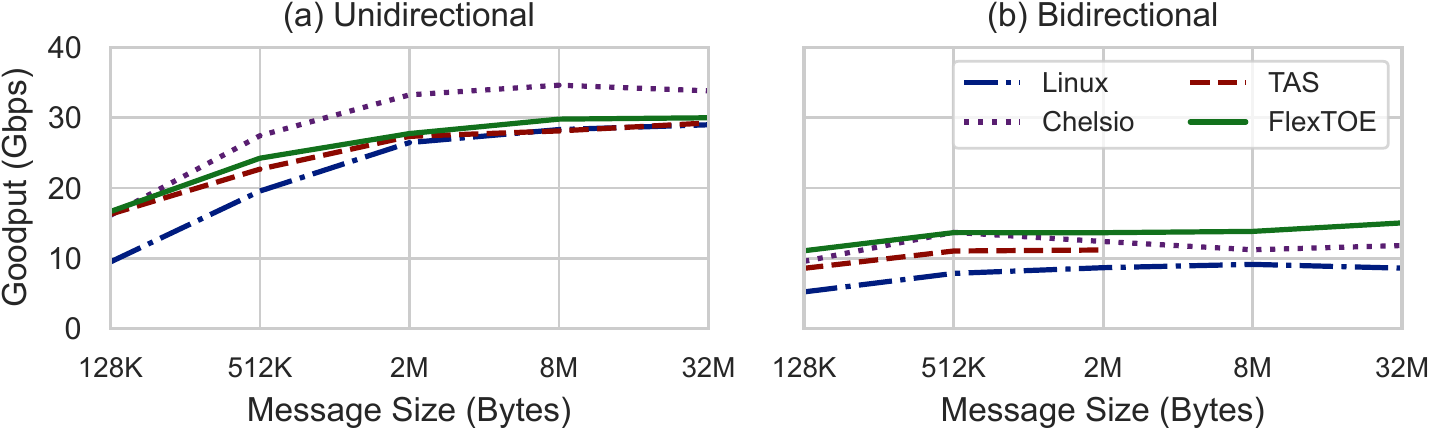}
  \caption{Large RPC throughput with varying RPC size.}
\label{fig:large-rpc}
\end{figure}

\paragraph{Per-connection throughput}
In this setup, a client transfers a large RPC message to the
server. In the first case (Figure~\ref{fig:large-rpc}a), the server
responds with a 32\,B response whereas in the second case (b), the
server echoes the message back to the client (TAS performance is
unstable with messages > 2\,MB in this case---we omit these results). In
the short-response case, Chelsio performs 20\% better than the other
stacks---Chelsio is a 100\,Gbps NIC optimized for unidirectional
streaming. However, it has 20\% lower throughput as compared to \sys
in the echo case. Other stacks cannot parallelize per-connection
processing, leading to limited throughput\footnote{With multiple
  unidirectional flows, all stacks achieve line rate
  (Figure~\ref{fig:packet-loss}b).}, while \sys's throughput is
limited by its protocol stage. \sys currently acknowledges every
incoming packet. For bidirectional flows, this quadruples the number
of packets processed per second. Implementing delayed ACKs would
improve \sys's performance further for large flows.

\begin{figure}
  \centering
  \includegraphics[width=\columnwidth]{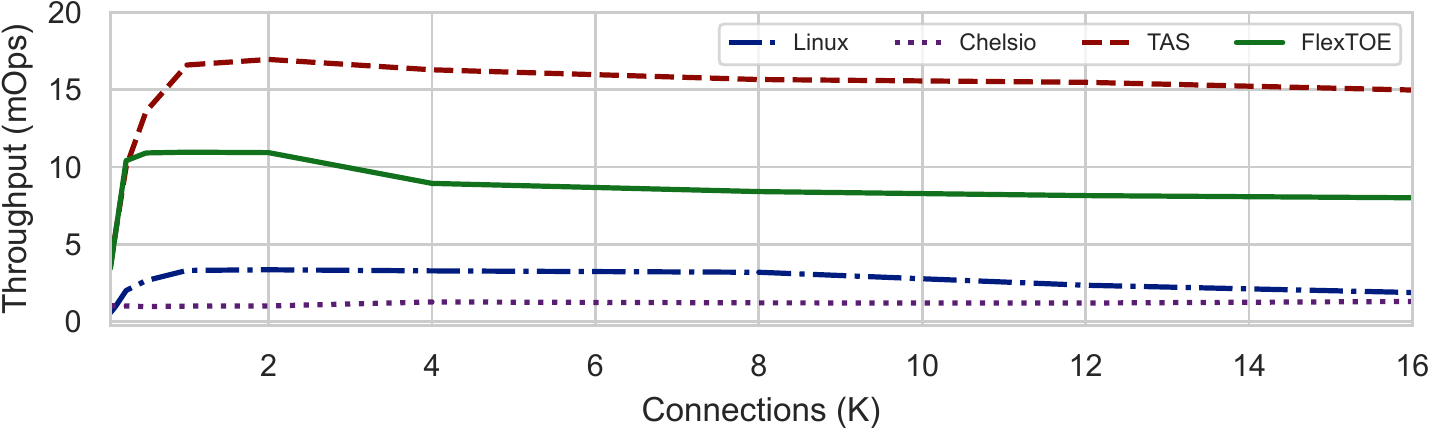}
  \caption{Connection scalability benchmark.}
\label{fig:conn-scale}
\end{figure}

\paragraph{Connection scalability} We establish an increasing number
of RPC client connections from all 5 client machines to a
multi-threaded echo server. To stress TCP processing, each connection
leaves a single 64\,B RPC in-flight. Figure~\ref{fig:conn-scale} shows
the throughput as we vary the number of connections. This workload is
very challenging for \sys as it exhausts fast memory and prevents
per-connection batching, causing a cache miss at every pipeline stage
for every segment. Up to 2K connections, \sys shows a throughput of
3.3$\times$ Linux. TAS performs 1.5$\times$ better than FlexTOE for
this workload. \sys is compute-bottlenecked\footnote{We expect that
  running \sys on the Agilio LX with 1.2\,GHz FPCs---1.5$\times$
  faster than Agilio CX---would boost the peak throughput to match TAS
  performance. Agilio LX also doubles the number of FPCs and
  islands. It would allow us to exploit more parallelism and cache
  more connections.} at the protocol stage, which uses 8 FPCs in this
benchmark. Agilio CX caches 2K connections in CLS memory. Beyond this,
the protocol stage must move state among local memory, CLS, and
EMEM. EMEM's SRAM cache is increasingly strained as the number of
connections increases. \sys's throughput declines by 24\% as we hit 8k
connections and plateaus beyond that\footnote{While we evaluate up to
  16K connections, \sys can leverage the 2\,GB on-board DRAM to scale
  to 1M+ connections.}. TAS's fast-path exhibits better
connection scalability, as it has access to the larger host CPU cache,
while Linux's throughput declines significantly. Chelsio has poor
performance for this workload,
as \texttt{epoll()} overhead dominates.

\paragraph{Benefit of data-path parallelism} To break down the impact
of \sys's data-parallel design on RPC performance, we repeat the echo
benchmark with 64 connections, with each connection leaving a single
2\,KB RPC in-flight (to be able to evaluate both intra and inter
connection parallelism). Table~\ref{tab:parallelism} shows the
performance impact as we progressively add data-path parallelism. Our
baseline runs the entire TCP processing to completion on the SmartNIC
before processing the next segment. Pipelining improves performance by
46$\times$ over the baseline. As we enable 8 threads on the FPCs
(2.25$\times$ gain), we hide the latency of memory operations and
improve FPC utilization. Next, we replicate the pre-processing and
post-processing stages, leveraging sequencing and reordering for
correctness, to extract 1.35$\times$ improvement and finally, with
four flow-group islands, we see a further 2$\times$ improvement. We
can see that each level of data-path parallelism is necessary,
improving RPC throughput and latency by up to 286$\times$.

\begin{table}
  \small
\begin{tabular}{lrrrr}
\multirow{2}{*}{\textbf{Design}} &
                                   \multicolumn{1}{c}{\textbf{Throughput}}
  & \multirow{2}{*}{$\times$} & \multicolumn{2}{c}{\textbf{Latency} (us)}            \\
                                 & \multicolumn{1}{r}{(Mbps)}              && \multicolumn{1}{r}{50p} & \multicolumn{1}{r}{99.99p} \\
\midrule
Baseline                & 79.32    & 1 & 1,179 & 6,929 \\
+ Pipelining            & 3,640.49  & 46 & 183   & 684   \\
+ Intra-FPC parallelism & 8,194.34  & 103 & 128   & 148   \\
+ Replicated pre/post   & 11,086.93 & 140 & 94    & 106   \\
+ Flow-group islands    & 22,684.69 & 286 & 46    & 58
\end{tabular}
\caption{\sys data-path parallelism breakdown.\vspace{-17pt}}
\label{tab:parallelism}
\end{table}

\begin{figure}
  \centering
  \includegraphics[width=\columnwidth]{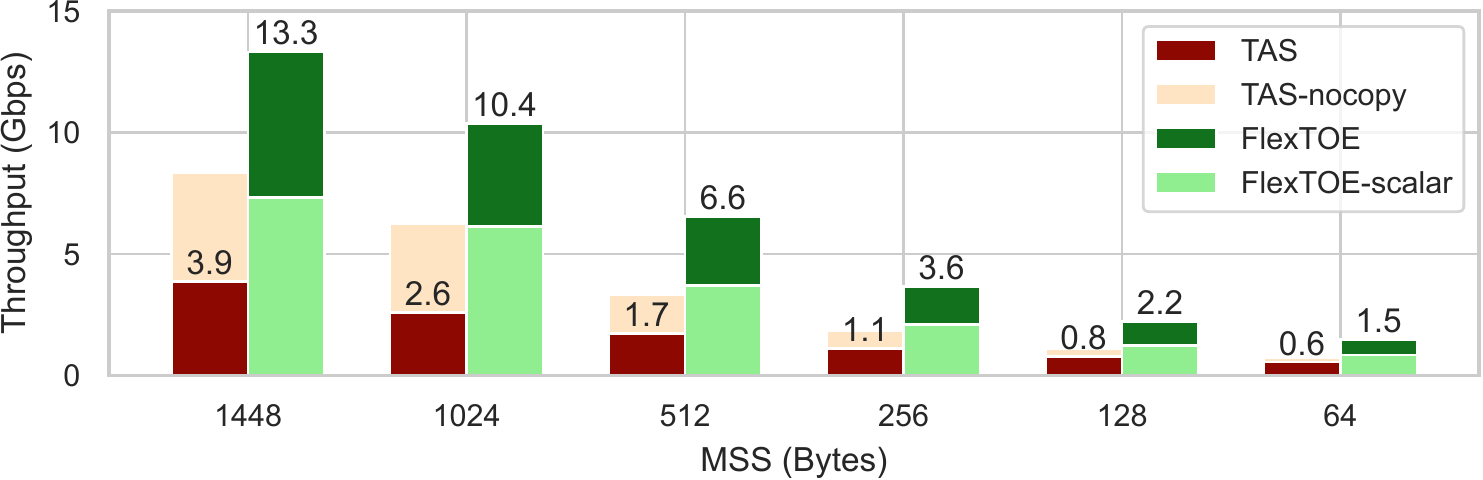}
  \caption{\sys benefits on BlueField SmartNIC.}
  \label{fig:bluefield}
\end{figure}

\paragraph{Do these benefits generalize?} We investigate whether
data-path parallelism provides benefits across platforms. In
particular, we investigate single connection throughput of pipelined
RPCs across a range of maximum segment sizes (MSS) on a Mellanox
BlueField~\cite{bluefield} MBF1M332A-ASCAT 25\,Gbps SmartNIC and on a
32-core AMD 7452 @ 2.35\,GHz host with 128\,GB RAM, 148\,MB aggregate
cache, and a conventional 100\,Gbps ConnectX-5 NIC.
We use a
single-threaded RPC sink application, running on the same
platform\footnote{BlueField is an off-path SmartNIC that is not
  optimized for packet processing offload to host-side applications
  (\S\ref{sec:nic_arch}).}. We compare TAS's core-per-connection
processing to FlexTOE's data-parallelism. We replicate each of \sys's
pre and post processing stages 2$\times$, resulting in 9 \sys
cores. Further gains may be achievable by more replication. To break
down \sys's benefits, we also compare to a \sys pipeline without
replicated stages (\sys-scalar), using 7 cores.

Figure~\ref{fig:bluefield} shows BlueField results. \sys outperforms
TAS by up to 4$\times$ on BlueField (and 2.4$\times$ on
x86). Depending on RPC size, \sys accelerates different stages of the
TCP data path. For large RPCs, \sys accelerates data copy to socket
payload buffers. To show this, we eliminate the step in TAS
(TAS-nocopy), allowing TAS to perform at 0.5$\times$ \sys on BlueField
(and identical to \sys on x86). For smaller RPCs, TAS-nocopy benefits
diminish and \sys supports processing higher packet rates. \sys-scalar
achieves only up to 2.3$\times$ speedup over TAS on BlueField (and
1.47$\times$ on x86), showing that only part of the benefit comes from
pipelining. Finally, \sys speedup is greater on the wimpier BlueField,
resembling our target architecture (\S\ref{sec:nic_arch}), than on
x86. To save powerful x86 cores, some stages may be collapsed, even
dynamically (cf.~Snap~\cite{snap}), at little performance cost.

\subsection{Robustness}\label{eval:loss}

\begin{figure}
  \centering
  \includegraphics[width=\columnwidth]{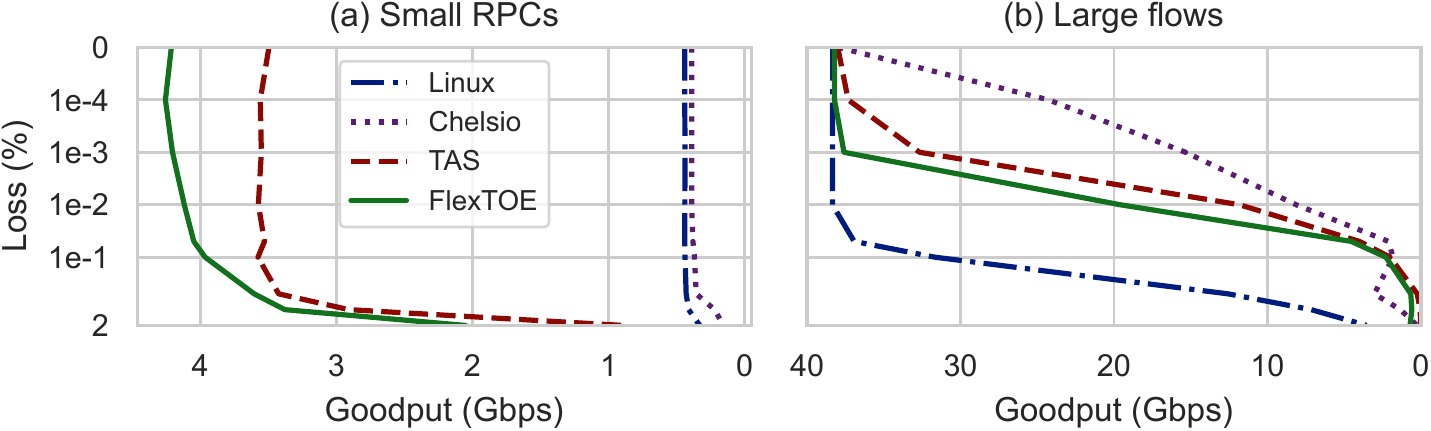}
  \caption{Throughput, varying packet loss rate.}
\label{fig:packet-loss}
\end{figure}

\paragraph{Packet loss} We artificially induce packet losses in the
network by randomly dropping packets at the switch with a fixed
probability. We measure the throughput between two machines for 100
flows running 64\,B echo-benchmark as we vary the loss probability,
shown in Figure~\ref{fig:packet-loss}a. We configure the clients to
pipeline up to 8 requests on each connection to trigger out-of-order
processing when packets are lost. \sys's throughput at 2\% losses is
at least twice as good as TAS and an order of magnitude better than
the other stacks for this case. We repeat the unidirectional large RPC
benchmark with 8 connections and measure the throughput as we increase
the packet loss rate. For this case (b), Chelsio has a very steep
decline in throughput even with $10^{-4} \%$ loss probability.  Linux
is able to withstand higher loss rates as it implements more
sophisticated reassembly and recovery algorithms, including selective
acknowledgments---\sys and TAS implement single out-of-order interval tracking
on the receiver-side and go-back-n recovery on the sender. \sys's
behavior under loss is still better than TAS. \sys processes
acknowledgments on the NIC, triggering retransmissions sooner, and its
predictable latency, even under load, helps \sys recover faster from
packet loss. We note that RDMA tolerates up to 0.1\%
losses~\cite{mittal2018revisiting}, while eRPC falters at 0.01\% loss
rate~\cite{erpc}. Unlike \sys, RDMA discards all out-of-order packets on the receiver
side~\cite{mittal2018revisiting}. TAS~\cite{tas} provides further
evaluation of the benefits of receiver out-of-order interval tracking.

\begin{figure}
  \centering
  \includegraphics[width=\columnwidth]{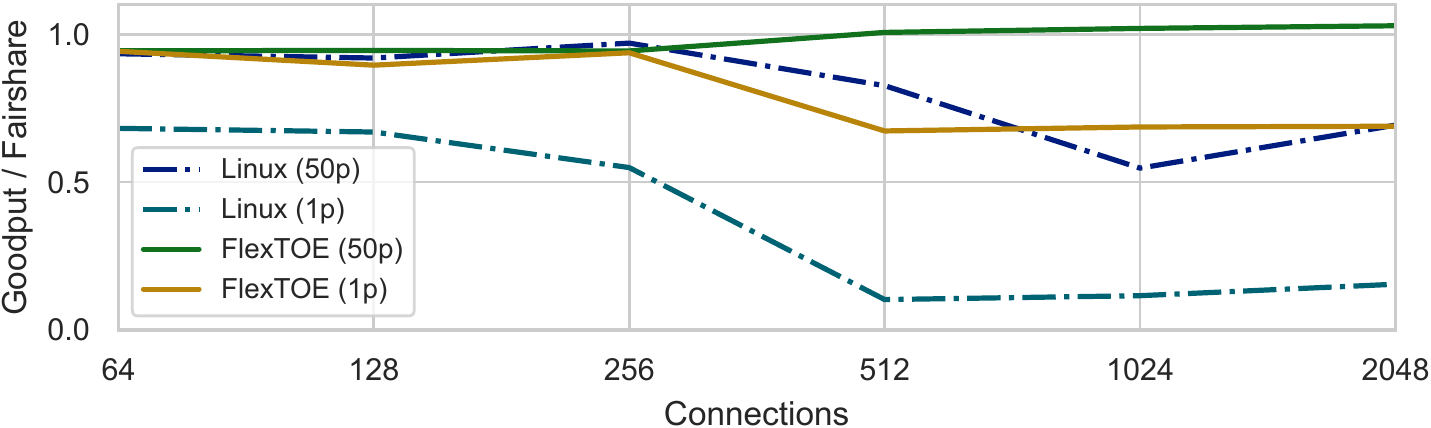}
  \caption{Throughput distribution at line rate.}
\label{fig:qmfair}
\end{figure}

\paragraph{Fairness} To show scalability of \sys's SCH
(\S\ref{sec:queueman}), we measure the distribution of connection
throughputs of bulk flows between two nodes at line rate for 60
seconds. Figure~\ref{fig:qmfair} shows the median and 1st percentile
throughput of \sys and Linux as we vary the number of connections. For
\sys, the median closely tracks the fair share throughput and the tail
is 0.67$\times$ of the median. Linux's fairness is significantly
affected beyond 256 connections. Jain's fairness index (JFI) drops to
0.36 at 2K connections for Linux, while \sys achieves 0.98. Above 1K
connections, Linux' median throughput is worse than \sys's 1st
percentile.

\paragraph{Incast} We simulate incast by enabling traffic shaping on
the switch to restrict port bandwidth to various incast degrees and we configure WRED to perform tail drops when the switch
buffer is exhausted. In this experiment, the client transfers 64\,KB
RPCs and the server responds with a 32\,B response on each
connection. As shown in Table~\ref{tab:incast}, control-plane-driven
congestion control in \sys is able to achieve the shaped line rate,
maintain low tail latency, and ensure fairness among flows under
congestion. Disabling it causes excessive drops, inflating tail latency by 18.8$\times$ and
skewing fairness by 2$\times$.

\begin{table}
\small
\begin{tabular}{lrrrrrrr}
\multirow{2}{*}{\textbf{deg.}} & \multirow{2}{*}{\textbf{\# con.}} & \multicolumn{2}{c}{\textbf{Tpt.} (G)} & \multicolumn{2}{c}{\textbf{Lat.} 99.99p (ms)} & \multicolumn{2}{c}{\textbf{JFI}} \\
& & on & off & on & off & on & off \\
\toprule
4 & 16 & 9.51 & 9.47 & 5.98 & 11.58 & 0.98 & 0.95 \\
4 & 64 & 9.51 & 9.23 & 10.75 & 44.39 & 0.96 & 0.73 \\
4 & 128 & 9.48 & 8.96 & 13.74 & 64.25 & 0.99 & 0.53 \\
10 & 10 & 3.66 & 1.04 & 2.50 & 18.26  & 0.95 & 0.78 \\
20 & 20 & 1.76 & 0.36 & 7.35 & 138.32 & 0.95 & 0.46 \\
\end{tabular}
\caption{\sys congestion control under incast.\vspace{-17pt}}
\label{tab:incast}
\end{table}
 \section{Conclusion}

\sys is a flexible, yet high-performance TCP offload engine to
SmartNICs. \sys leverages fine-grained parallelization of the TCP
data-path and segment reordering for high performance on wimpy
SmartNIC architecture, while remaining flexible via a modular
design. We compare \sys to Linux, the TAS software TCP accelerator,
and the Chelsio Terminator TOE. We find that Memcached scales up to
38\% better on \sys versus TAS, while saving up to 81\% host CPU
cycles versus Chelsio. \sys provides competitive performance for RPCs,
even with wimpy SmartNICs, and is robust under adverse operating
conditions. \sys's API supports XDP programs written in eBPF. It allows us
to implement popular data center transport features, such as TCP
tracing, packet filtering and capture, VLAN stripping, flow
classification, firewalling, and connection splicing.

\paragraph{Acknowledgments.} We thank the anonymous reviewers and our
shepherd, Brent Stephens, for their helpful comments and
feedback. This work was supported by NSF grant 1751231.

\bibliographystyle{plain}
\bibliography{paper}

\clearpage
\appendix

\section{TCP Connection State Partitioning}\label{app:state}

To enable fine-grained parallelism, we partition connection state
across pipeline stages. Table~\ref{tab:flowstate} shows the
per-connection state variables, grouped by pipeline
stage. Pre-processor state contains connection identifiers (MAC, IP
addresses; TCP port numbers). Protocol state contains TCP windows,
sequence and acknowledgment numbers, and host payload buffer
positions. Post-processor state contains host payload buffer and
context queue locations, and data-path congestion control state. DMA
and context queue stages are stateless.

In aggregate, each TCP connection has 108 bytes of state, allowing us
to offload millions of connections to the SmartNIC. In particular, we
can manage 16 connections per protocol FPC, 512 connections per
flow-group, and 16K connections in the EMEM cache. Using all of EMEM,
we can support up to 8M connections.

\begin{table}
\begin{tabular}{lrl}
Field & Bits & Description\\
    \midrule
    \multicolumn{3}{l}{\textbf{Pre-processor} (connection
    identification)---15B:}\\
    \texttt{peer\_mac} & 48 & Remote MAC address \\
    \texttt{peer\_ip} & 32 & Remote IP address \\
    \texttt{local|remote\_port} & 32 & TCP ports \\
    \texttt{flow\_group} & 2 & hash(4-tuple) \% 4\\
    \multicolumn{3}{l}{\textbf{Protocol} (TCP state machine)---43B:}\\
    \texttt{rx|tx\_pos} & 64 & RX/TX buffer head\\
    \texttt{tx\_avail} & 32 & Bytes ready for TX\\
    \texttt{rx\_avail} & 32 & Available RX buffer space\\
    \texttt{remote\_win} & 16 & Remote receive window\\
    \texttt{tx\_sent} & 32 & Sent unack. TX bytes\\
    \texttt{seq} & 32 & TCP seq. number\\
    \texttt{ack} & 32 & TCP remote seq. number\\
    \texttt{ooo\_start|len} & 64 & Out-of-order interval\\
    \texttt{dupack\_cnt} & 4 & Duplicate ACK count\\
    \texttt{next\_ts} & 32 & Peer timestamp to echo\\
    \multicolumn{3}{l}{\textbf{Post-processor} (ctx queue, congestion control)---51B:}\\
    \texttt{opaque} & 64 &  App connection id\\
    \texttt{context} & 16 & Context-queue id\\
    \texttt{rx|tx\_base} & 128 & RX/TX buffer base\\
    \texttt{rx|tx\_size} & 64 & RX/TX buffer size\\
    \texttt{cnt\_ackb|ecnb} & 64 & ACK’d and ECN bytes\\
    \texttt{cnt\_fretx} & 8 & Fast-retransmits count\\
    \texttt{rtt\_est} & 32 & RTT estimate\\
    \texttt{rate} & 32 & TX rate \\
\end{tabular}
  \caption{Connection state partitions (total: 108B).\vspace{-15pt}}
  \label{tab:flowstate}
\end{table}

\section{Connection Splicing Implementation}

We implement AccelTCP's connection splicing in 24 lines of eBPF
code. Listing~\ref{lst:splicexdp} shows the entire code.

\begin{listing}[ht]
\begin{minted}
[frame=lines,
fontsize=\footnotesize,
escapeinside=||,
]{C}
|\textcolor{heraldBlue}{BPF\_MAP\_HASH\_DECLARE}|(splice_tbl, SPLICE_MAX_FLOWS, \
    sizeof(struct pkt_4tuple_t), sizeof(struct tcp_splice_t));

int bpf_xdp_prog(struct |\textcolor{heraldBlue}{xdp\_md}|* ctx)
{
  struct tcp_splice_t state;
  struct pkt_hdr_t     *hdr = |\textcolor{heraldBlue}{BPF\_XDP\_ADDR}|(ctx->data);
  struct pkt_4tuple_t *key = &hdr->ip.src;

  // Filter non-IPv4/TCP segments to control-plane
  if (!segment_ipv4_tcp(hdr))
    return |\textcolor{heraldRed}{XDP\_REDIRECT}|;

  // Connection Control: Segments with SYN, FIN, RST
  // Atomically remove map entry and forward to control-plane
  if (segment_tcp_ctrlflags(hdr)) {
    |\textcolor{heraldBlue}{BPF\_MAP\_DELETE\_ELEM}|(splice_tbl, key);
    return |\textcolor{heraldRed}{XDP\_REDIRECT}|;
  }

  if (|\textcolor{heraldBlue}{BPF\_MAP\_LOOKUP\_ELEM}|(splice_tbl, key, &state) < 0)
    return |\textcolor{heraldRed}{XDP\_PASS}|;    // Send to data-plane

  patch_headers(hdr, &state);
  return |\textcolor{heraldRed}{XDP\_TX}|;    // Send out the MAC
}

void patch_headers(struct pkt_hdr_t *hdr,
                   struct tcp_splice_t *state)
{
  hdr->eth.src   = hdr->eth.dst;
  hdr->eth.dst   = state->remote_mac;
  hdr->ip.src    = hdr->ip.dst;
  hdr->ip.dst    = state->remote_ip;
  hdr->tcp.sport = state->local_port;
  hdr->tcp.dport = state->remote_port;

  hdr->tcp.seq  += state->seq_delta;
  hdr->tcp.ack  += state->ack_delta;
}
\end{minted}
\caption{Connection splicing with XDP in \sys.}
\label{lst:splicexdp}
\end{listing}

\section{TAS TCP/IP Processing Breakdown}\label{sec:tas_breakdown}

\begin{table}
\begin{tabular}{lrr}
   \textbf{Function}          & \textbf{Cycles} & \textbf{\%} \\ \toprule
   Segment generation         & 130             & 9           \\
   Loss detection (and recovery) & 606             & 42          \\
   Payload transfer           & 10              & 1          \\
   Application notification   & 381             & 26          \\
   Flow scheduling            & 172             & 12          \\
     Miscellaneous              & 141             & 10 \\
     \hline
     Total & 1,440 & 100
   \end{tabular}
   \caption{Breakdown of TCP/IP stack overheads in TAS.\vspace{-15pt}}
   \label{tab:tcp-overheads}
\end{table}

Table~\ref{tab:tcp-overheads} shows a breakdown of the per-packet
TCP/IP processing overheads (summarized as \emph{TCP/IP stack} in
Table~\ref{tab:overheads}) in TAS for the Memcached benchmark
conducted in \S\ref{sec:tcp_overhead}. For each request, TAS performs
loss detection (and potentially recovery) that involves processing the
incoming request segment, generating an acknowledgement for it, and
additionally, processing the acknowledgement for the response segment,
consuming 42\% of the total per-packet processing cycles.  TAS spends
9\% of the total cycles to prepare the response TCP segment for
transmission and an additional 12\% to schedule flows based on the
rate configured by the congestion control protocol. TAS spends 26\% of
per-packet cycles interacting with the application, to notify when a
request is received, to admit a response for transmission, and to free
the transmission buffer when it is acknowledged. For small
request-response pairs (32B in this case), the payload copy overheads
are negligible.

\section{Control Plane}\label{sec:controlplane}

\sys's control plane is similar to that of existing approaches that
separate control and data-plane activities, such as
TAS~\cite{tas}. Using it, we implement control-plane policies, such as
congestion control, per-connection rate limits, per-application
connection limits, and port partitioning among applications
(cf.~\cite{kernelinterposeflexible}). We briefly describe connection
and congestion control in this appendix. Retransmissions are described
in \S\ref{sec:hcflow} and \S\ref{sec:rxflow}. TAS~\cite{tas} provides
further description and evaluation of the control plane (named
``slow-path'' in the TAS paper).

\paragraph{Connection control} Connection control involves complex
control logic, such as ARP resolution, port and buffer allocation, and
the TCP connection state machine. The data-path forwards control
segments to the control-plane. The control-plane notifies \libsys of
incoming connections on listening ports. If the application decides to
\texttt{accept()} the connection, the control-plane finishes the TCP
handshake, allocates host payload buffers and a unique connection
index for the data-path. It then sets up connection state in the
data-path at the index location. Similarly, \libsys forwards
\texttt{connect()} calls to the control-plane, which establishes the
connection. On \texttt{shutdown()}, the control-plane disables the
connection and removes the corresponding data-path state.

\paragraph{Congestion control} \sys provides a generic control-plane
framework to implement different rate and window-based congestion
control algorithms, akin to that in TAS~\cite{tas}. The control-plane
runs a loop over the set of active flows to compute a new transmission
rate, periodically. The interval between each iteration of the loop is
determined by the round-trip time (RTT) of each flow. In each
iteration, the control-plane reads per-flow congestion control
statistics from the data-path to calculate a new rate or window for
the flow. The rate or window is then set in the data-path flow
scheduler (\S\ref{sec:queueman}) for enforcement. We also monitor
retransmission timeouts in the control iteration. \sys implements
DCTCP~\cite{dctcp} and TIMELY~\cite{timely} in this way.

\section{\sys x86 and BlueField Ports}\label{sec:ports}

We have ported the \sys data-path to the x86 and BlueField
platforms. \sys's design across the different ports is identical. We
do not merge or split any of the fine-grained modules or reorganize
the pipeline across ports. \sys's decomposition, pipeline parallelism,
and per-stage replication all generalize across platforms. Both ports
are also almost identical to the Agilio-CX40 implementation
(cf.~\S\ref{sec:impl}) and were completed within roughly 2
person-weeks, demonstrating the great
development velocity of a software TCP offload engine. We describe the
implementation differences of each port to the Agilio-CX40 version in
this section.

\paragraph{Hardware cache management.} The hardware-managed cache
hierarchies of x86 and BlueField obviate the need for software-managed
caching that was implemented on Agilio. Instead of leveraging
near-memory processing acceleration of the NFP-4000
(cf.~\S\ref{sec:near_memory}), our ports implement multi-core ring
buffers, flow lookup and packet sequencers in software. The more
powerful x86 and BlueField cores make up for the difference in
performance.

\paragraph{Symmetric core mapping.} Unlike the NFP-4000, where FPCs
are organized into islands, cores on x86 and BlueField have mostly
symmetric communication properties, so the assignment of modules to
cores is arbitrary and the manual FPC mapping step is
omitted. However, we note that core mapping may still be beneficial,
for example to leverage shared caches and node locality on
multi-socket x86 systems. Each instance of a module runs on
its own core. Apart from the six fine-grained pipeline modules:
\emph{pre-processing}, \emph{protocol}, \emph{post-processing}, \emph{DMA},
\emph{context queue}, and \emph{SCH} shown in Figure \ref{fig:pipeline},
the ports utilize an additional \emph{netif} module to interface with
DPDK NIC queues to receive and transmit packets. Therefore, FlexTOE-scalar
uses 7 cores and the FlexTOE-2$\times$ configuration uses 2 additional
cores to replicate the pre and post-processing stages for a
total of 9 cores.

\paragraph{Context queues use only shared memory.} Our x86 and
BlueField ports currently only support applications running on the
same platform as \sys. Hence, context queues always use shared memory
rather than DMA. The corresponding DMA pipeline stage executes the
payload copies in software using shared memory, rather than leveraging
a DMA engine.

\paragraph{Platform-specific parameters.} The replication factor of
each pipeline stage is platform dependent. Stage-specific
microbenchmarks on each platform can determine it. Our generalization
experiments (\S\ref{eval:rpc}) are designed to show that \sys's
data-parallelism can improve single connection throughput. Hence, we
configure only one instance of the \sys data-path pipeline in these
versions (no flow-group islands---we do not process multiple
connections in these experiments). Each port's pipeline uses the same
number of stages as the Agilio-CX40 version, but we set different
replication factors for the pre and post processing stages on x86 and
BlueField (no replication and 2$\times$ replication). We do not
attempt to find the optimal replication factor for best performance
nor compact stages to reduce wasted CPU cycles.

\end{document}